\documentclass[biblatex,nocompat]{trbunofficial}

\usepackage{subfigure}
\usepackage{graphicx}

\usepackage[hidelinks]{hyperref}

\usepackage[utf8]{inputenc}

\AuthorHeaders{Krishnakumari, Cats, and van Lint}
\title{Day-to-day and seasonal regularity of network passenger delay for metro networks}

\author{%
  \textbf{Panchamy Krishnakumari}\\
  Department of Transportation and Planning\\
Delft University of Technology\\
P.O. Box 5048, 2600 GA Delft, The Netherlands\\  
Phone number: +31 15 27 85279 \\
Email: p.k.krishnakumari@tudelft.nl
\\
  \hfill\break
  \textbf{Oded Cats}\\
  Department of Transportation and Planning\\
Delft University of Technology\\
P.O. Box 5048, 2600 GA Delft, The Netherlands  \\
Phone number: +31 15 27 85279 \\
Email: o.cats@tudelft.nl\\
  \hfill\break%
  \textbf{Hans van Lint}\\
Department of Transportation and Planning \\
Delft University of Technology \\
P.O. Box 5048, 2600 GA Delft, The Netherlands  \\
Phone number: +31 15 27 85279 \\
Email: j.w.c.vanlint@tudelft.nl
}

\begin{document}
\maketitle

\section{Abstract}

 In an effort to improve user satisfaction and transit image,  transit service providers worldwide offer delay compensations. Smart card data enables the estimation of passenger delays throughout the network and aid in monitoring service performance. Notwithstanding, in order to prioritize measures for improving service reliability and hence reducing passenger delays, it is paramount to identify the system components - stations and track segments - where most passenger delay occurs. To this end, we propose a novel method for estimating network passenger delay from individual trajectories. We decompose the delay along a passenger trajectory into its corresponding track segment delay, initial waiting time and transfer delay. We distinguish between two different types of passenger delay in relation to the public transit network: average passenger delay and total passenger delay. We employ temporal clustering on these two quantities to reveal daily and seasonal regularity in delay patterns of the transit network. The estimation and clustering methods are demonstrated on one year of data from Washington metro network. The data consists of schedule information and smart card data which includes passenger-train assignment of the metro network for the months of August 2017 to August 2018. Our findings show that the average passenger delay is relatively stable throughout the day. The temporal clustering reveals pronounced and recurrent and thus predictable daily and weekly patterns with distinct characteristics for certain months. 

\hfill\break%
\noindent\textit{Keywords}: passenger delay, estimation, seasonal patterns, daily patterns, passenger-train assignment, smart card data
\newpage
\section{Introduction}
\label{sec:intro}

Service reliability is known to be one of the most important determinants of performance, ridership and user satisfaction. In an effort to improve user satisfaction and transit image, several transit authorities have introduced policies for offering delay compensations. Such measures include refunds by Dutch Railways (NS) in The Netherlands \cite{ns}, by SL in Stockholm, Sweden \cite{sl} and the so-called rush hour promise by WMATA in Washington DC, USA \cite{wmata}. In order to provide such services, collecting, monitoring and quantifying the state of public transportation systems in terms of for example delays, occupancy, regularity and punctuality is critically important. The data with which many of these quantities can be directly or indirectly estimated are already collected by many transport authorities through different sensors and information sources. Some of the well-known and increasingly used data sources for public transport network are infrastructure (track segments) and service network (line) information, time table data (schedule information), automatic vehicle location (AVL) data which contains real time locations of trains \cite{moreira2015improving} and by implication the realisation of the schedule, and automatic fare collection (AFC) or smart card data with origin-destination specific information of passengers \cite{pelletier2011smart}. 

AFC systems typically come in two forms: tap-in only or tap-in and tap-out. Tap-in only fare systems registers the boarding station of the passenger only whereas tap-in-tap-out systems collect both origin and destination locations of each passenger. Notwithstanding, there are methods to infer the alighting station of each individual using tap-in only AFC system \cite{trepanier2007individual, sanchez2017inference}. An increasing body of literature is available on the use of smart card data in public transit; a review of which can be found in \cite{pelletier2011smart,koutsopoulos2017automated}. Most of these studies focus on origin-destination matrix estimation \cite{nassir2011transit,gordon2013automated}; extracting passenger patterns \cite{ma2013mining,bhaskar2014passenger,ma2017understanding}; network performance analysis \cite{ma2015modeling,ma2017quantile} and passenger-train assignment \cite{zhao2016estimation,zhu2017probabilistic}. 

Using the readily available timetable and AVL data, it is easy to estimate the vehicle delay of the individual trains. However, for many applications it is more interesting and fruitful to investigate the \textit{passenger delay} for a given line, network segment or transfer station. Clearly, with the increasing availability of AFC data and progress in passenger-train assignment research, it is possible to estimate and quantify this delay for each passenger. A passenger-train assignment model is necessary here, since AFC systems do not collect information about the actual trajectory of each passenger. Passenger delay derived from AFC data is origin-destination specific for each passenger and hence is valid for a given realised trajectory only. To attribute all these individual person delays to delay statistics of individual public transport network elements (routes, lines, nodes) over different periods of time, we need to aggregate and average the individual passenger delays of all passengers traversing through the network on each network element.

In this work, we study passenger delay as extract from fusing AVL and AFC data as the service performance indicator and infer it per network element. This is the key contribution of this paper: a new data-driven method to derive PT network delays from individual trajectories. The methodology has similarities with the data driven OD estimation method we propose in \cite{Krishnakumari2019od}, in that we construct a solvable system of equations utilising all the information at hand without making more assumptions than strictly needed. Usage of passenger-train assignment data make this study unique because this type of information is relatively new. To the authors' knowledge, this is the first study to explore the potential application of this unique data set. In this work, we distinguish two different types of passenger delay in relation to the public transit network: average passenger delay and total passenger delay. We define the \textit{average passenger delay} as the delay incurred by a passenger while traversing a track segment (link), station (node) or trajectory. The \textit{total passenger delay} is the total delay experienced by all the passengers that traverse that link, node or route. Thus, the total passenger delay is a function of the number of passengers that traverse that network element during a given time period.

In the remainder of this paper we show that realised passenger-trajectories (resulting from AFC data and passenger-train assignment) and schedule information are sufficient to estimate average and total passenger delays for all the different network elements. The estimation improves by adding an additional constraint derived from AVL data, because these are---in our case---readily available. With these constraints for each passenger and each vehicle, a solvable system of equations can be formulated. There are various applications for the resulting network indicators, such as modeling delay propagation though the network; disruption detection, etc. In this work, we use these estimated delay patterns to reveal day-to-day or seasonal regularity by grouping the delay of different days. We demonstrate the framework for the metro network of Washington DC with one year of data.

The paper is organized as follows: section \ref{sec:method} describes the overall estimation framework and the clustering to reveal the day-to-day and seasonal regularity of the delay patterns; in section \ref{sec:data} we outline the network and data used. These results are presented in section \ref{sec:results}. We offer conclusions and a discussion on further research avenues in section \ref{sec:conclusion}.
\section{Methodology}
\label{sec:method}

In this section we first define the notations used in this work. Thereafter, we describe the network passenger delay estimation problem as a solvable set of equations and finally, we elaborate the clustering technique used to reveal the regularities in the delay patterns.

Let $G(S,E,L,V^{trans})$ denote a directed graph representing a public transport network, in which $S$ is the set of stations; $E$ the set of track segments between the stations; $L$ the set of ordered pairs of stations representing a public transport service line; and $V^{trans}$ (subset of $S$) the set of transfer stations where transfers between lines occur. Passenger trajectory or route $r_{s_o,s_d,n}$ between origin station $s_o$ and destination station $s_d$ of passenger $n \in <1,...N>$, where $N$ is the total number of passengers, is defined using two sets, one of stops and one of lines, each of which are a subset of $S$ and $L$ respectively. The combination of $S$ and $L$ allows one to define a third set: the set of track segments (subset of $E$) that is traversed by the passenger. As a result, we obtain the initial stop (first element in the stop set); the intermediate transfer stops (second to one before last element in the stop set) and the link set. The stop set is defined as $<s_1, .., s_i, s_{i+1}, .., s_m>$, where $s_2, .., s_{i}, .. s_{m-1}$ are the $m$ transfer stations for that route, in which $s_1$ is the origin station $s_o$ and $s_m$ is the destination station $s_d$. The line set for the same route is defined as $<l_1, .., l_i, .. l_{m-1}>$.

\subsection{Network delay estimation}
Assuming that the AFC system is a tap-in-tap-out system with the passenger-train assignment provided; the origin and destination station, and the inferred trajectory of each passenger are assumed known. Thus, the passenger delay $d_{s_o,s_d,k,n}$ is defined as:

\begin{equation}\label{pax_delay}
    d_{s_o,s_d,k,n} = \begin{cases}
    t_{s_o,s_d,k,n} - \widetilde{t}_{s_o,s_d,k} & if~t_{s_o,s_d,k,n} > \widetilde{t}_{s_o,s_d,k} \\
    0 & otherwise
    \end{cases}
\end{equation}

where $t_{s_o,s_d,k,n}$ is the observed travel time for passenger $n$ departing at time (period) $k$ that can be obtained by finding the difference between the tap-in and tap-out time of that passenger. Any negative delay $d_{s_o,s_d,k,n}$ is assigned to $0$ as shown in eq.\ref{pax_delay} since this implies the passenger reached the destination earlier than expected. The maximum scheduled route time $\widetilde{t}_{s_o,s_d,k}$ can be defined in several ways. In this work, the following definition has been adopted:
\begin{equation}
\label{eq:1}
    \widetilde{t}_{s_o,s_d,k} = \sum_{l \in r_{s_o,s_d}}\sum_{s_i \in l}\widetilde{t}_{s_i,s_{i+1},k}^{train} + \sum_{l \in r_{s_o,s_d}}{h_{l,k}} + \sum_{s_i \in r_{s_o,s_d}}\sum_{s_i=s_1}^{s_{m-1}}{t_{s_i}^{walk}}
\end{equation}
where $\widetilde{t}_{s_o,s_d,k}$ is composed of scheduled running times and dwell times $\widetilde{t}_{s_i,s_{i+1},k}^{train}$ assigned to track segments connecting subsequent stations $s_i$ and $s_{i+1}$, headway $h_{l,k}$ between successive train services for the lines $l$ in that route and walking time $t_{s_i}^{walk}$ at the origin station and transfer stations denoted by $s_i$. The headway between trains are included in calculating the schedule travel time due to the definition of on-time journey considered in this study. A passenger is considered to be on-time with regards to the maximum scheduled route time even he/she just missed a train but was able to catch the next train of that line and the headway between the trains account for that fallback time. 

Given the individual trajectories of the passengers, the aim is to decompose the delay experienced by the passengers to the corresponding network elements they traversed along their route. We assume that the observed travel time of a passenger $t_{s_o,s_d,k,n}$ comprises of travel time that can be attributed to one of the following network elements of the public transport infrastructure: 
\begin{enumerate}
\item Time spent at the origin stop of the trip 
\item On-board a train along one of the trip segments
\item Time spent at the transfer stations of the trip
\end{enumerate}

\begin{figure}[h!]
    \centering
    \includegraphics[scale=0.4]{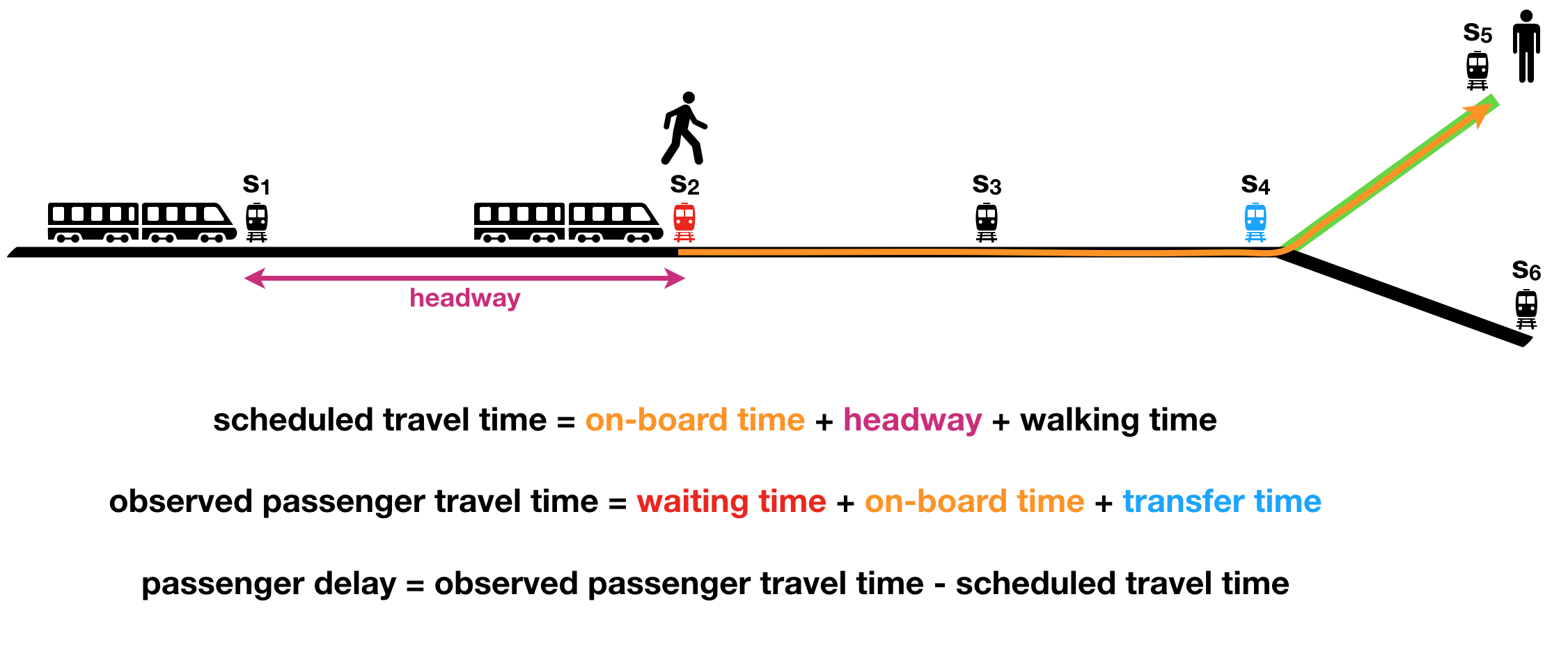}
    \caption{Schematic picture of scheduled and observed travel time between station $s_2$ and $s_5$.}
    \label{fig:toy}
\end{figure}

Similarly, the delay experienced by a passenger is comprised of the delays occurring at these network elements. The composition of the scheduled and observed passenger travel time is illustrated in figure~\ref{fig:toy}. Based on this schematic representation, the passenger delay between stations $(s_o,s_d)$ for a given departure time $k$ can be defined as:

\begin{equation}\label{pax_delay_exp}
    d_{s_o,s_d,k,n} \geq d_{s_o,k}^{wait} + \sum_{l \in r_{s_o,s_d,n}}\sum_{s_i \in l}d_{s_i,s_{i+1},k}^{on-board} + \sum_{s_i \in r_{s_o,s_d,n}}\sum_{s_i = s_1}^{s_{m-1}}d_{s_i,k}^{trans}
\end{equation}

where $d_{s_o,k}^{wait}$ is the initial waiting delay at the origin station $s_o$, $d_{s_i,s_{i+1},k}^{on-board}$ is the on-board delay between track segment $(s_i,s_{i+1})$ and $d_{s_i,k}^{trans}$ is the transfer delay at transfer stop $s_i$ in route $r_{s_o,s_d,n}$. The inequality in eq.\ref{pax_delay_exp} is due to potential \textit{non-observable} personal travel components which may add additional delays such as random passenger arrival time or performing an activity at one of the stations within the gated area. Formulating the relationship between personal delays and the three network related delay components by means of an inequality constraint allows us to perform the delay inference without ignoring unobserved personal delay components. With eq.\ref{pax_delay_exp}, now each passenger trajectory can be written as a linear combination of three passenger delay component types. Doing so for all passenger trajectories leads to a potentially solvable system of equations (inequalities), since each passenger’s trip between $k$ and $k + t_{s_o,s_d,k,n}$ serves as a constraint for all other trips that traverse one or more common network elements during this trip. The number of unknowns corresponds to the number of initial stations, transfer stations and links (connections between successive stations) in the public transport system. 

There is, however, an additional constraint we can formulate, that ---as it turns out---indeed renders the resulting system of equations  solvable. This constraint pertains to the on-board delay component, which must be equal to the train delay between the corresponding track segments. This train delay can be directly inferred from the train movement data and the schedule information. The result is an additional constraint to the system of equations in eq.\ref{pax_delay_exp} which reads
\begin{equation}\label{train_delay_exp}
    d_{s_o,s_d,k}^{train} = \sum_{l \in r_{s_o,s_d}}\sum_{s_i \in l}d_{s_i,s_{i+1},k}^{on-board},
\end{equation}
where $d_{s_o,s_d,k}^{train}$ is the train delay for a train servicing the line that starts at $s_o$ and ends at $s_d$ within time $k$, $r_{s_o,s_d}$ is the route composed of track segments visited by the train and the initial waiting time and transfer delay component is 0 as those are not relevant for a train. 

Eq.\ref{pax_delay_exp} and eq.\ref{train_delay_exp} now constitute a solvable system of equations in which $d_{s_o,k}^{wait}~\forall~s_o \in S$, $d_{s_i,s_{i+1},k}^{on-board}~\forall~(s_i,s_{i+1}) \in E$ and $d_{s_i,k}^{trans}~\forall~s_i \in V^{trans}$ are the unknowns and $d_{s_o,s_d,k,n}$ and $d_{s_o,s_d,k}^{train}$ are known. This system of equations is solved using a constrained linear least square solution method \cite{altman1999regularized} with a lower bound set to 0 to ensure a non-negative solution.

\subsection{Temporal clustering}

Now that the passenger delay has been decoupled into corresponding delays at the various network elements (segments, nodes), we can use these values to compute network performance indicators that are indicative for the state of a public transit network at any given time. We can define these states for different time periods / days and use these representations for example to explore if there is seasonal or daily regularity in the patterns. 

To represent the network state (dynamics), we use average passenger delay and total passenger delay. The average passenger delay equals the estimated delays ($d_{s_o,k}^{wait}$, $d_{s_i,s_{i+1},k}^{on-board}$ and $d_{s_i,k}^{trans}$) for each corresponding network element respectively, whereas the total passenger delay equals the average passenger delay weighted by the number of passengers that experienced that delay. Thus, the average passenger delay of a given day composed of the three components is represented using the following feature vector:
\begin{equation}
\label{eq:fv1}
    FV^{average} = [d_{s_o,k}^{wait},~ d_{s_i,s_{i+1},k}^{on-board},~ d_{s_i,k}^{trans}]
\end{equation}
The total passenger delay is represented as follows:
\begin{equation}
\label{eq:fv2}
        FV^{total} = [d_{s_o,k}^{wait}\times q_{s_o,k}^{wait},~ d_{s_i,s_{i+1},k}^{on-board}\times q_{s_i,s_{i+1},k}^{on-board},~ d_{s_i,k}^{trans}\times q_{s_i,k}^{trans}]
\end{equation}
where $q_{s_o,k}^{wait}$, $q_{s_i,s_{i+1},k}^{on-board}$ and $q_{s_i,k}^{trans}$ are the number of passengers that experienced initial waiting time, on-board and transfer delay respectively.

For studying the dynamics of the passenger delay, we use the vectorized form of the feature vectors in eq.\ref{eq:fv1} and eq.\ref{eq:fv2}. In order to avoid biasing the clustering from small and large outlier delay values, we normalised the feature vectors by clustering the delay values into uniform clusters such that track segments with 0 to 5 mins delays belong to cluster 1 and those with delays of 5 to 10 mins belong to cluster 2, if the aggregation is 5 mins. Thus the feature vector is composed of cluster number instead of delay values but each cluster has a corresponding known range of delay values.

The aim of the clustering is to study if regular patterns emerge for different days. These insights can be used for developing demand-oriented operational planning for different days or different months. In this work, we use hierarchical agglomerative clustering for the temporal clustering \cite{rokach2005clustering}. This is because of the power of such clustering in revealing the distribution of the feature vectors in the form of a dendrogram \cite{everittcambridge} which can aid in determining the optimal number of clusters as shown in \cite{krishnakumari2018understanding}. The hierarchy or distribution of the feature vectors is constructed based on a dissimilarity metric between the feature vectors. The dissimilarity metric used in this work is the city block distance.

For each cluster, we build a representative feature vector by taking the centroid of the feature vectors belonging to that class. Thus, for each class we have a stacked delay distribution plot of the three network element delay of the centroid in order to make the analysis informative and comprehensible. The distribution curve of centroid of each class $c$ for average passenger delay is defined as:
\begin{equation}
    dist_{c,k} = \sum_{s_o \in S}d_{s_o,k}^{wait}(c) : \sum_{s_i \in S}\sum_{s_{i+1} \in S}d_{s_i,s_{i+1},k}^{on-board}(c) : \sum_{s_i \in S}d_{s_i,k}^{trans}(c)
\end{equation}
where $d_{s_o,k}^{wait}(c)$, $d_{s_i,s_{i+1},k}^{on-board}(c)$ and $d_{s_i,k}^{trans}(c)$ represents the waiting time, on-board delay and transfer delay of the centroid of class $c$. $dist_{c,k}$ represents the ratio between all delay components in that cluster $c$ at time $k$. Similarly, a stacked distribution curve for the total passenger delay  is built. 
\section{Data}
\label{sec:data}
\begin{figure}[h!]
    \centering
    \includegraphics[scale=0.45]{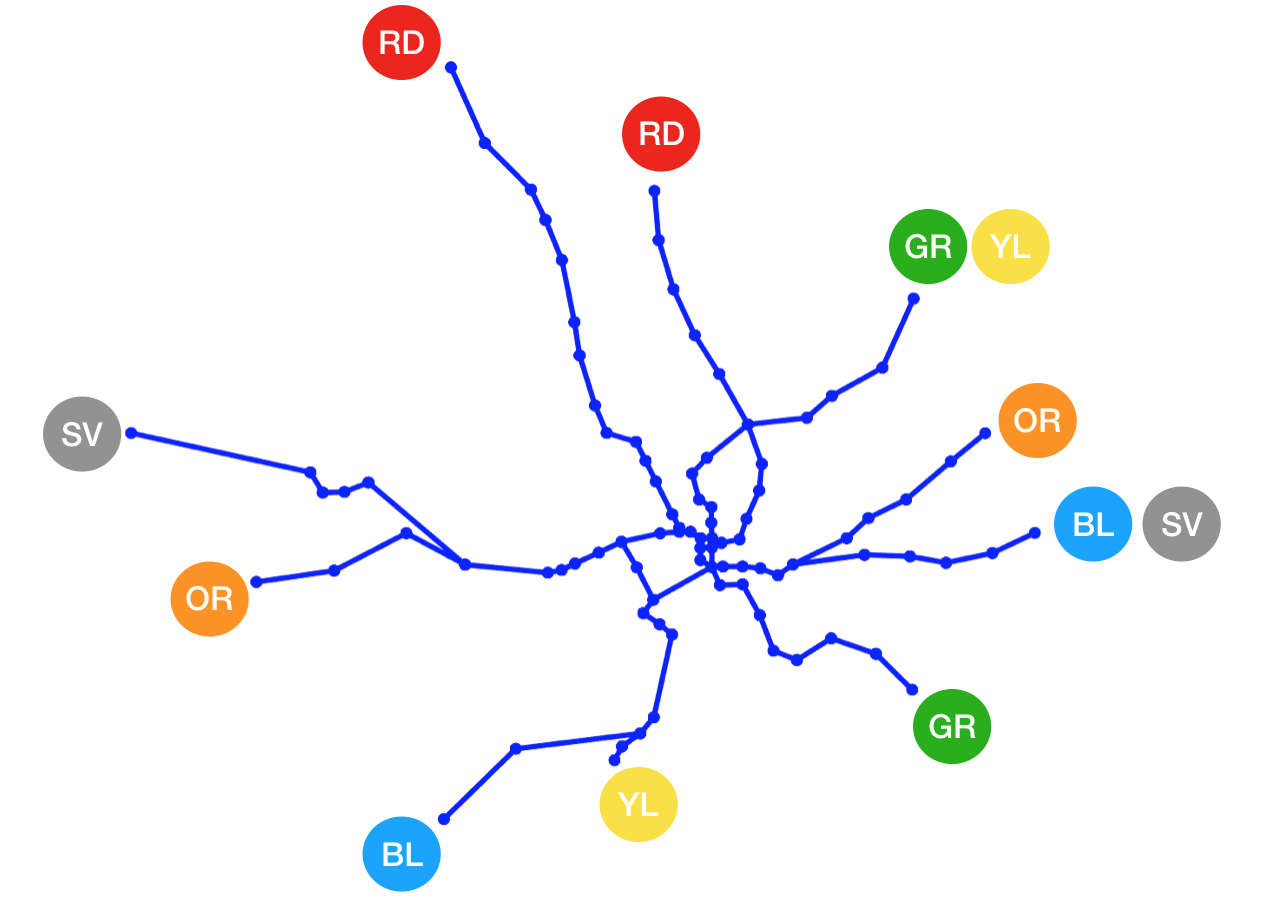}
    \caption{Washington metro network}
    \label{fig:network}
\end{figure}

We demonstrate the estimation and clustering approach on a data set of smart card data from WMATA for the region of Washington DC, USA. The data is composed of one year of smart card data from 19 August 2017 to 28 August 2018 from the whole metro network of Washington DC which includes the passenger-train assignment outputs derived from an application of the so-called ODX method described in \cite{sanchez2017inference}. The data also includes the rail movement data and schedule information. The metro network is comprised of 6 lines, 91 stations, 186 links and 9 transfer stations as shown in Figure~\ref{fig:network}.

The network has an average ridership of $\approx$ 438 000 rides per day and a total of $\approx$ 157 million rides during the entire study period with an average travel time of 28 minutes per passenger. 14\% of the passengers experience a mean delay of 6 minutes or longer. 39\% of the passenger trips include transfers with an average of 1.14 transfers per passenger trip. 
\section{Results}
\label{sec:results}
\begin{figure}[h!]
\centering  
\subfigure[]{\includegraphics[width=0.45\linewidth]{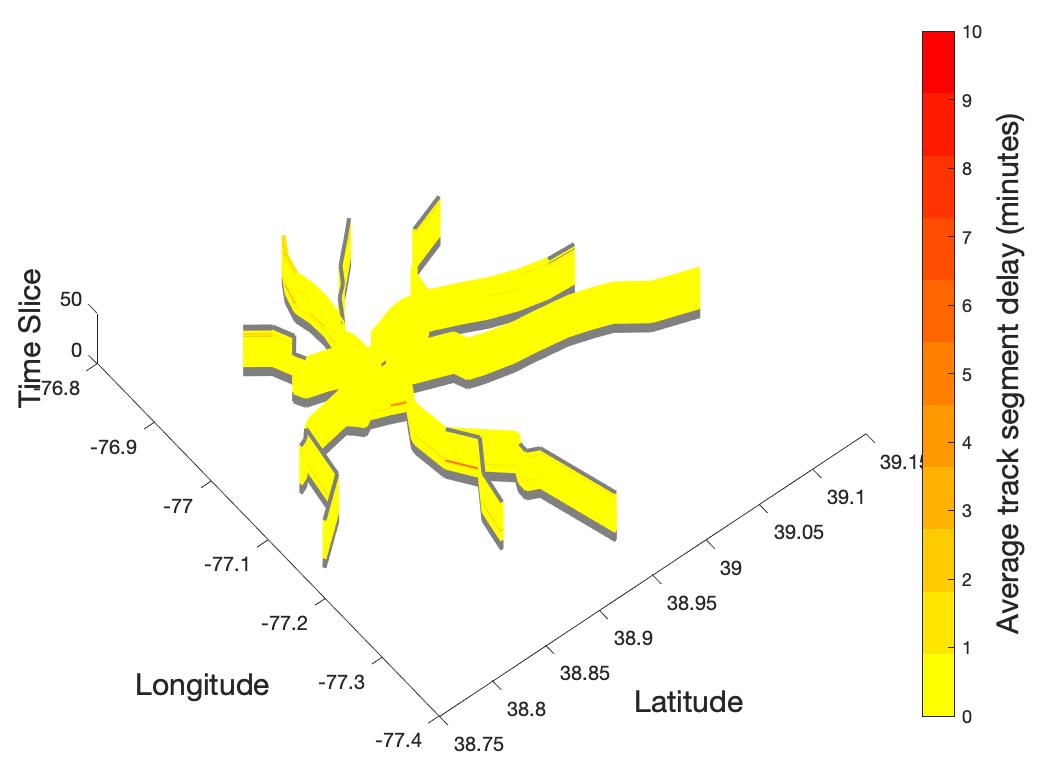}}
\subfigure[]{\includegraphics[width=0.45\linewidth]{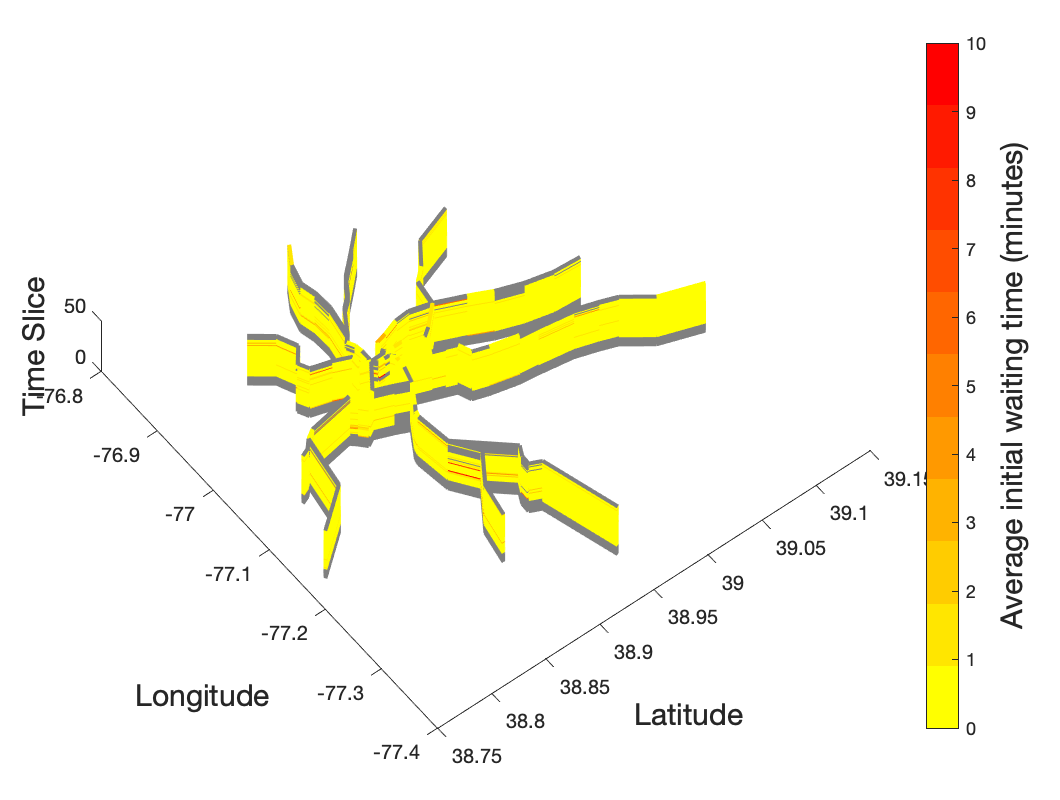}}
\subfigure[]{\includegraphics[width=0.45\linewidth]{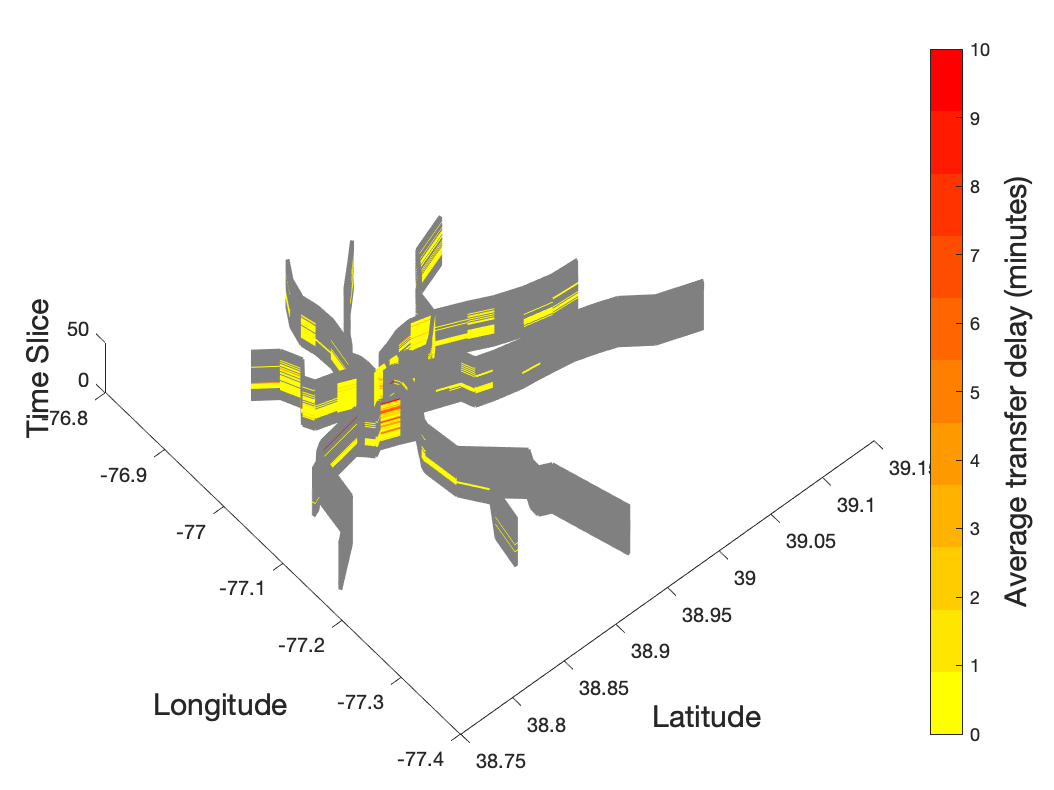}}
\caption{Average passenger delay estimation results for 01-09-2017 (a) Track segment delay (b) Initial waiting time (c) Transfer delay}
\label{fig1}
\end{figure}

In this section, we present the results of the delay estimation and the temporal clustering for the Washington metro network. We chose a temporal aggregation of 30 minutes since the maximum headway between trains is 20 minutes and choosing an aggregation less than that would imply that there would be no trains between some OD pairs, hence no passengers and consequently no system of equations. Having an aggregation of 30 minutes ensures that at least one train  per time slice is included in the system of equations as represented in eq.\ref{pax_delay_exp}. Figure~\ref{fig1} shows the estimation results of the average passenger delay for the three network elements for a selected day with an temporal aggregation of 30 minutes.

From figure~\ref{fig1} it can be seen that there is no significant link delay for this particular day. The waiting time and transfer delay is mapped on a link rather than a node as we incorporate directionality of the node to distinguish journeys between different lines. Thus, we can build these three compact 3D graph for visualising the delay propagation of each day. This can be used to evaluate the performance of the metro network at the end of the day or estimate the passenger delay incurred between any given origin-destination (OD) pair.

\begin{figure}[h!]
\centering  
\subfigure[]{\includegraphics[width=0.49\linewidth]{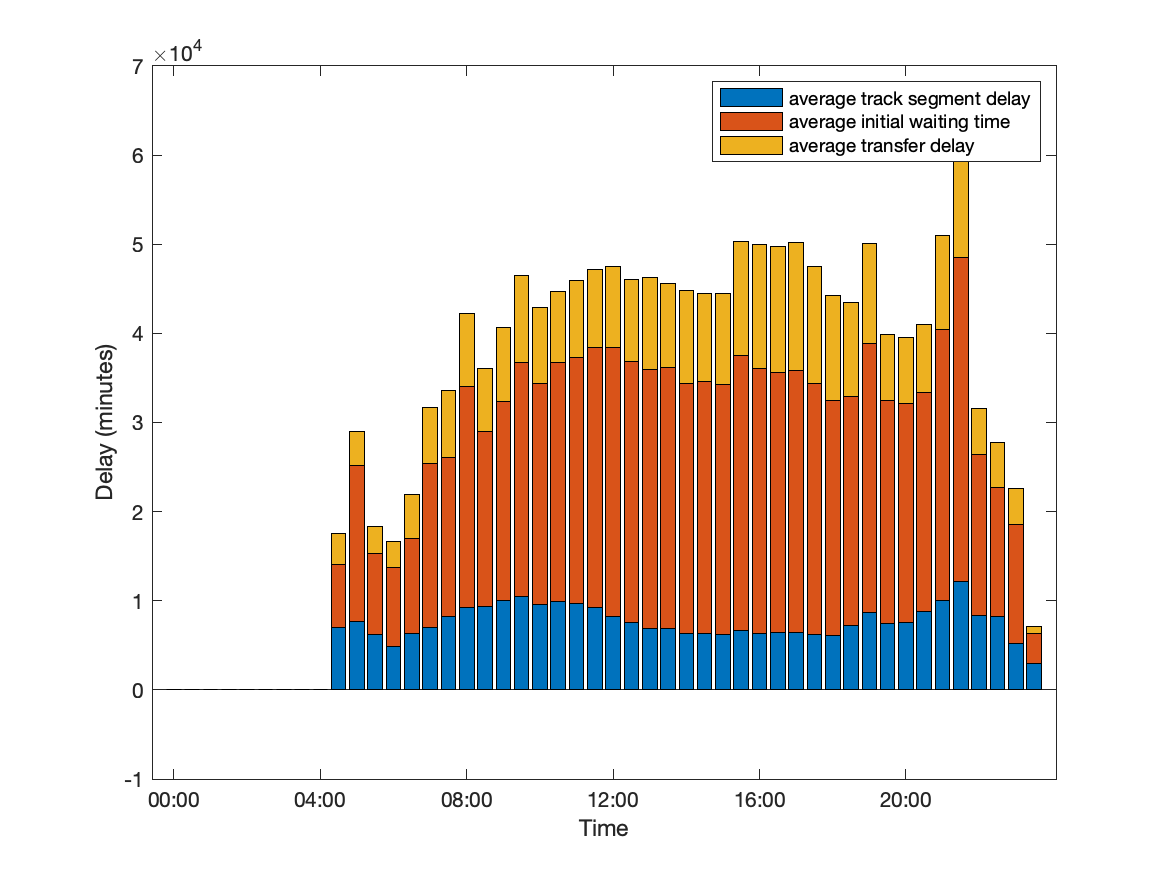}}
\subfigure[]{\includegraphics[width=0.49\linewidth]{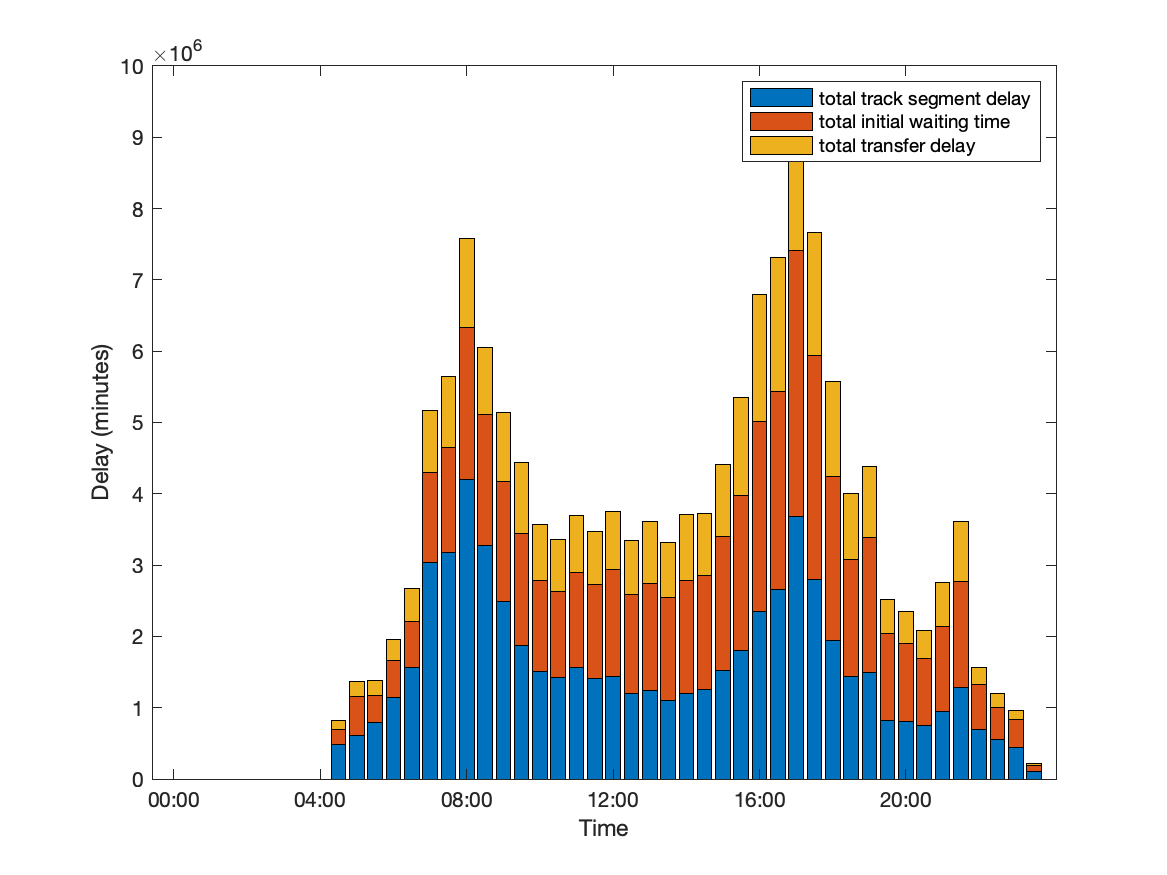}}
\subfigure[]{\includegraphics[width=0.47\linewidth]{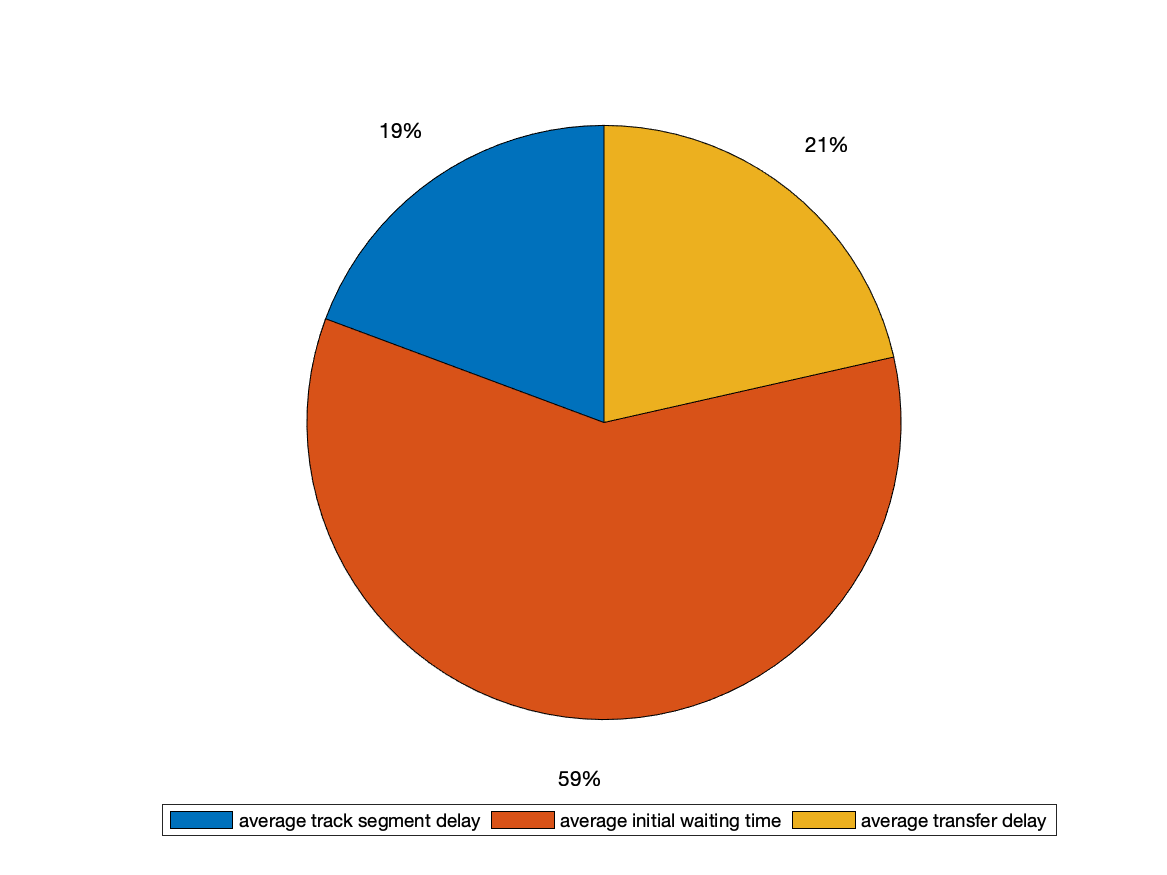}}
\subfigure[]{\includegraphics[width=0.49\linewidth]{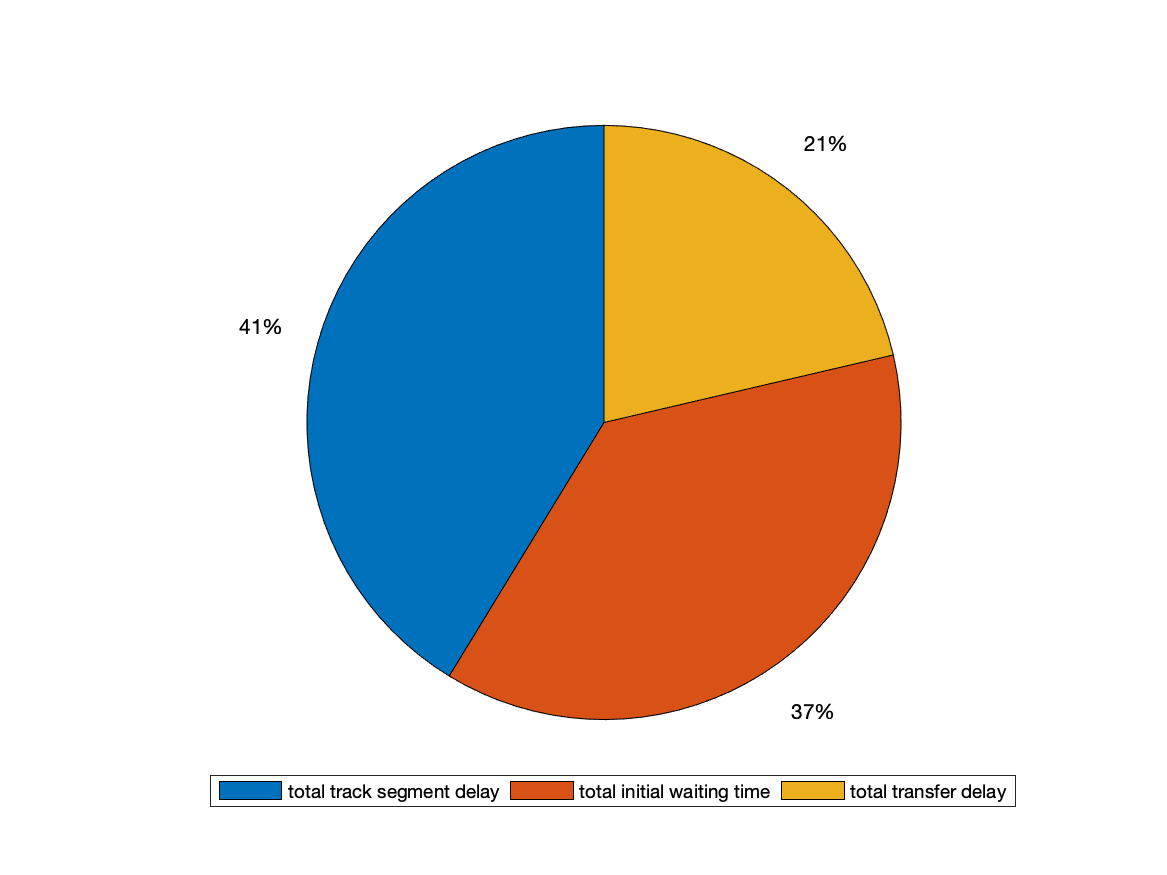}}
\caption{Delay distribution of the whole dataset across the three network elements for different time periods based on (a) average passenger delay and (b) total passenger delay.  Overall delay distribution of the whole dataset across the three network elements based on (c) average passenger delay and (d) total passenger delay}
\label{fig2}
\end{figure}

A look at the overall distribution of the average and total passenger delays for the entire analysis period decomposed into different time periods is shown in figure~\ref{fig2}. The average passenger delay, shown in figure~\ref{fig2}(a), is stable compared to the morning and evening peak in the total passenger delay distribution in figure~\ref{fig2}(b). Thus, delays occur at all time periods, more passengers are affected during the peak periods as can be expected. Figures~\ref{fig2}(c) and (d) shows the contribution of each network element delay to the overall delay. In the case of average passenger delay, 59\% of the delay is associated with the initial waiting time but when the number of passengers affected is considered, track segment delay contributes the most with 41\% of the delay. Thus, link delay is more correlated with flows than initial waiting times.


\begin{figure}[h!]
    \centering
    \includegraphics[scale=0.5]{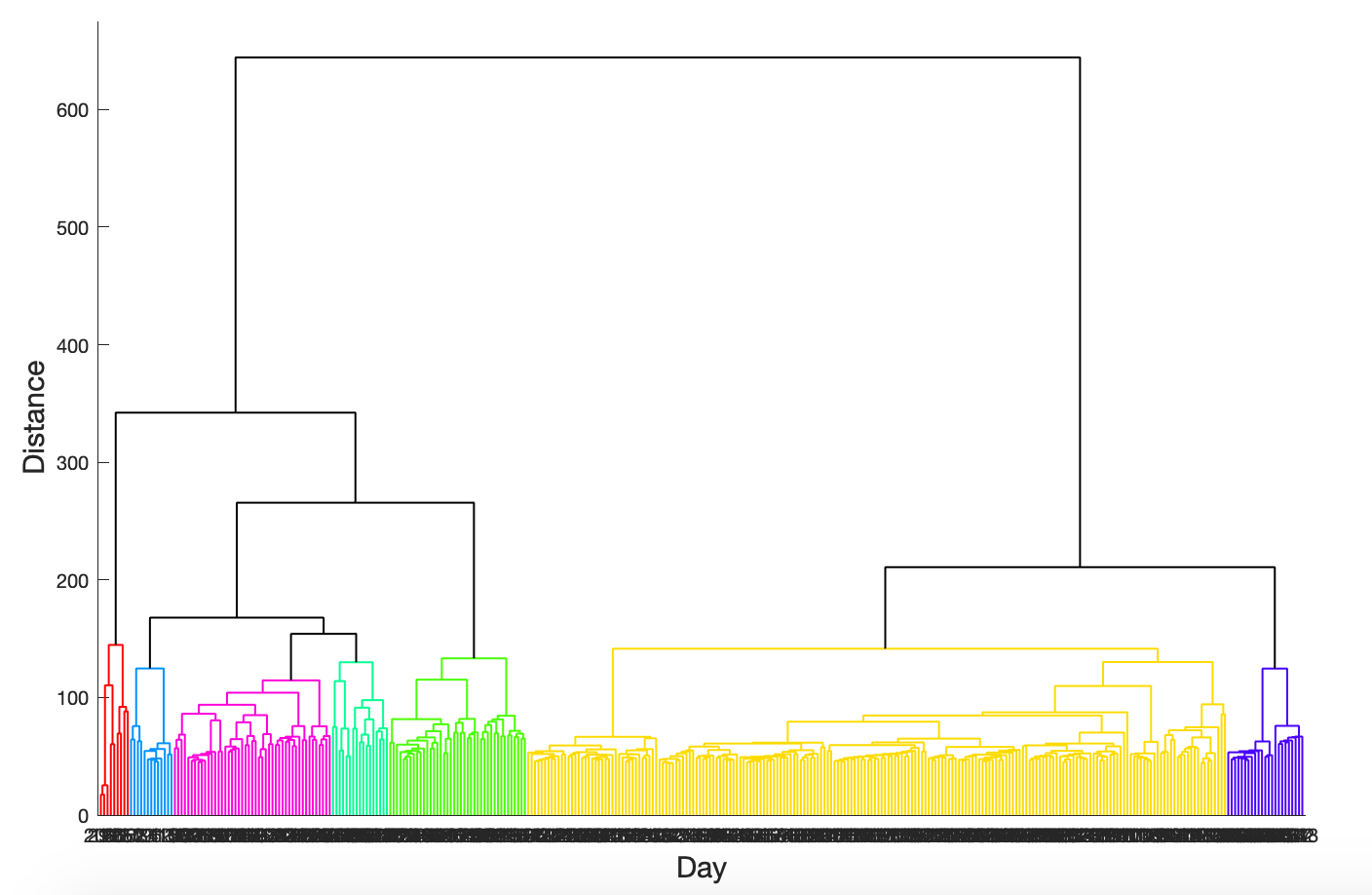}
    \caption{Dendrogram of the temporal clustering of average passenger delay}
    \label{fig:average}
\end{figure}

The estimated delays for the three network elements are clustered into uniform groups which are used to construct the feature vectors and temporally clustered. The results of the temporal clustering are analysed using a dendrogram for either different days or different months to extract day-to-day and seasonal variances respectively. Figure~\ref{fig:average} shows the dendrogram of the temporal clustering for average passenger delay. The number of leaves in the dendrogram corresponds to the number of days included in the dataset. There are clearly well-defined clusters with similar days clustered together. We chose the number of clusters to be 7 as it is a comprehensible number of clusters as well as we wanted to investigate if setting the number of cluster to 7 would extract distinct weekday or weekend patterns with clusters consisting only of Mondays, Tuesdays, etc. We can easily merge or further divide the clusters based on the insights we gain from these clusters.

The average passenger delay classes in the dendrogram in figure~\ref{fig:average} are numbered from left to right from 1 to 7 where the distance of 1 represents a delay range of 0 to 10 minutes. As can be observed the size of the clusters obtained varies greatly, from 9 to 209 days. The largest class consists of almost 60\% of the data. However, within this large cluster, the distance between the feature vectors varies from 60 to 120 - implying total passenger delay differences of up to 600 to 1200 minutes. This means that within a cluster, some feature vectors have high dissimilarity index whereas others are more similar.

\begin{figure}[h!]
    \centering
    \includegraphics[scale=0.5]{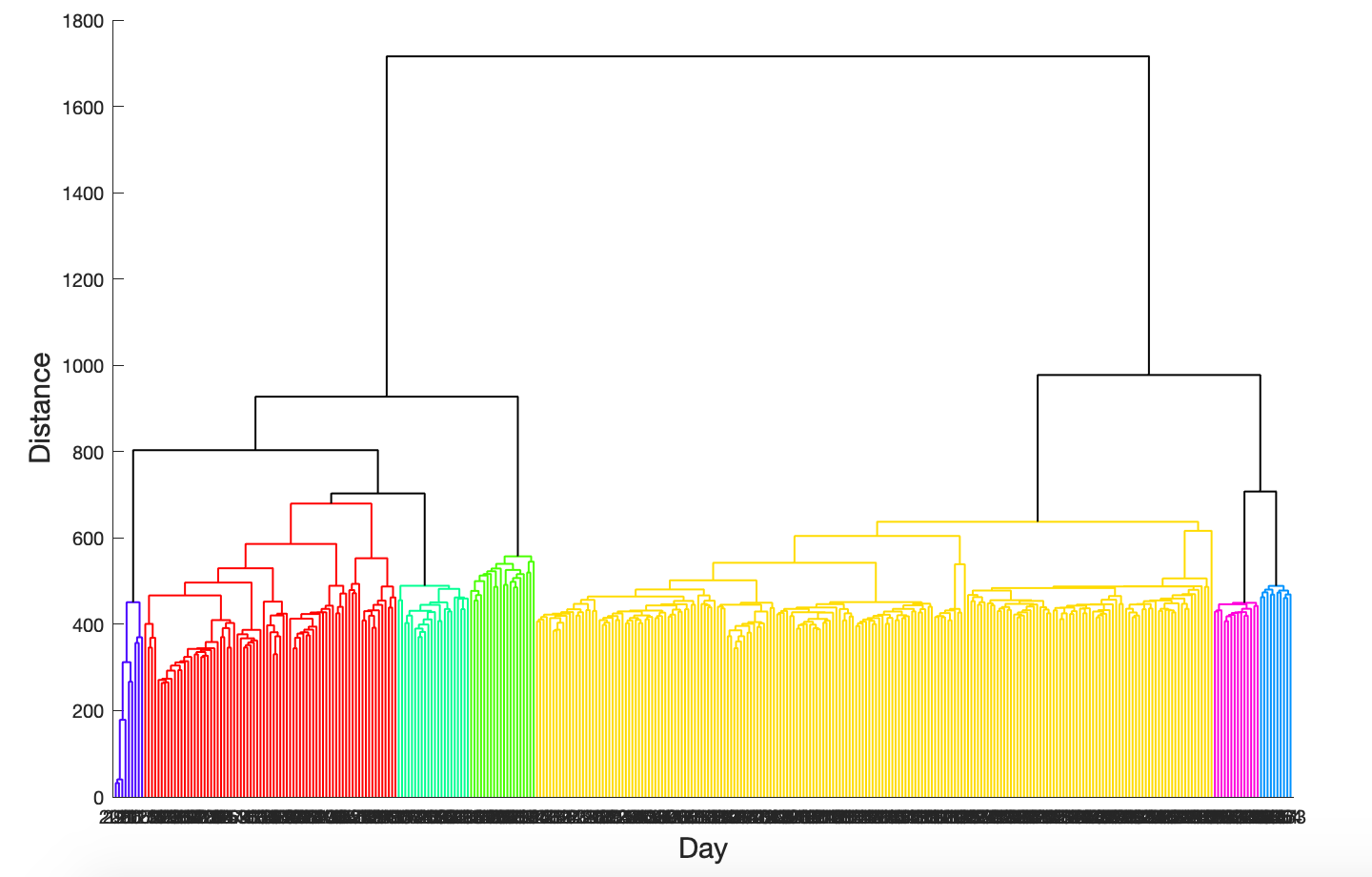}
    \caption{Dendrogram of the temporal clustering of total passenger delay}
    \label{fig:total}
\end{figure}

Similar to the temporal clustering of the average passenger delay, the total passenger delay can also be clustered. The resulting dendrogram is shown in Figure~\ref{fig:total}. We chose the number of cluster as 7 here as well to further analyse the data within each individual cluster. The total passenger delay classes in the dendrogram in figure~\ref{fig:total} are numbered from left to right from 1 to 7.

\begin{table}[h!]
\centering
\caption{Description of network states based on average passenger delay.}
\label{classes1}
\begin{tabular} {|c|c|c|c|}
\hline
\textbf{Class} &
\textbf{Delay distribution}  & 
\textbf{Distribution of months} & 
\textbf{Distribution of days}\\ 
\hline
1 &
 \begin{minipage}{.3\textwidth}
 \vspace*{0.02in}
  \includegraphics[trim = 20 0 0 0, width=\linewidth, height=29mm]{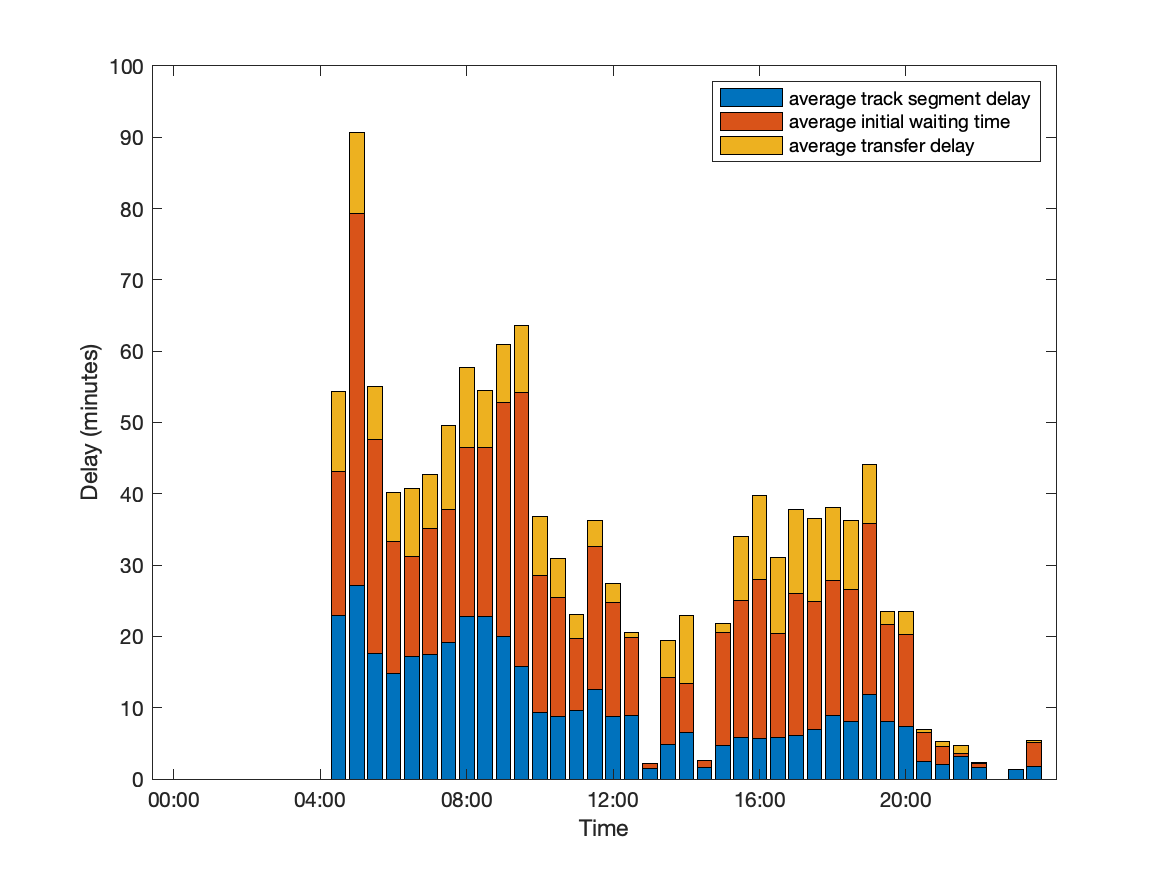}
    \end{minipage} & 
 \begin{minipage}{.25\textwidth}
 \vspace*{0.02in}      \includegraphics[trim = 20 0 0 0, width=\linewidth, height=29mm]{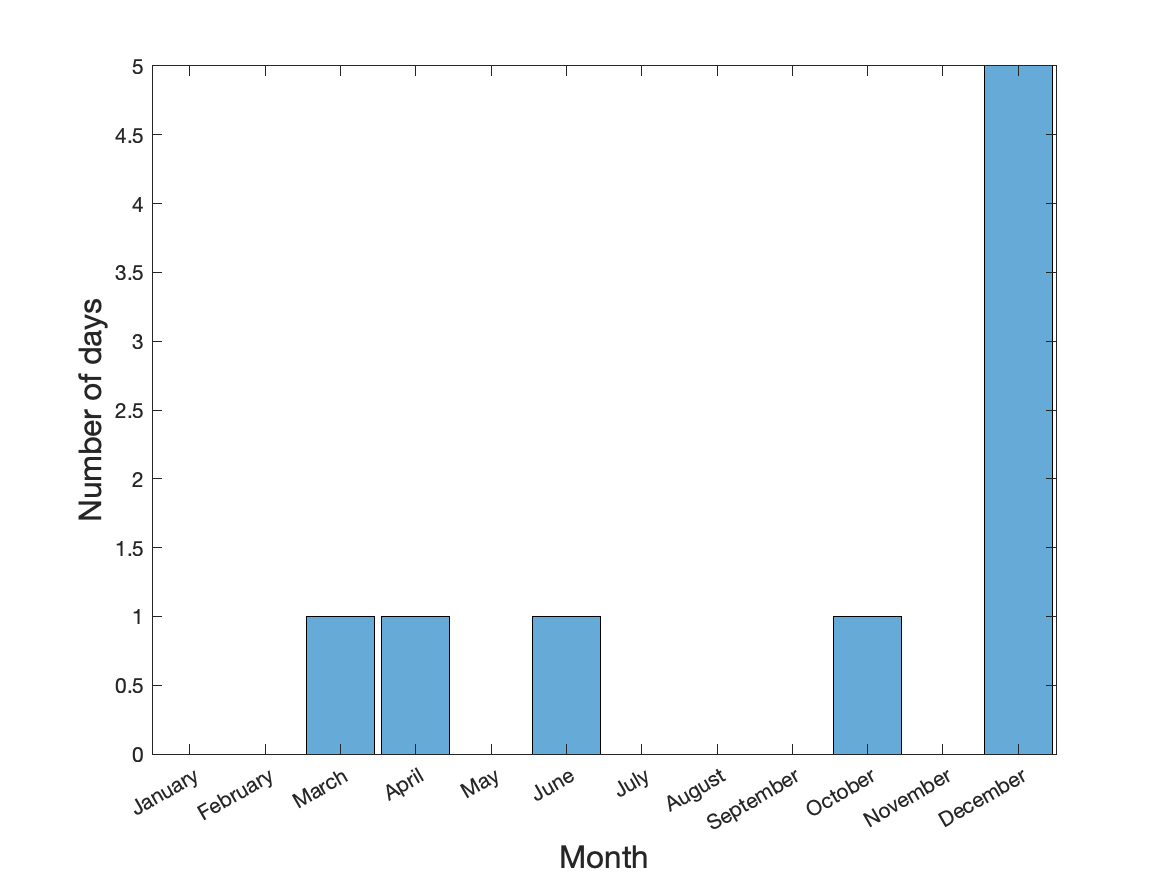}
    \end{minipage}  & 
 \begin{minipage}{.25\textwidth}
 \vspace*{0.02in}   \includegraphics[trim = 20 0 0 0, width=\linewidth, height=29mm]{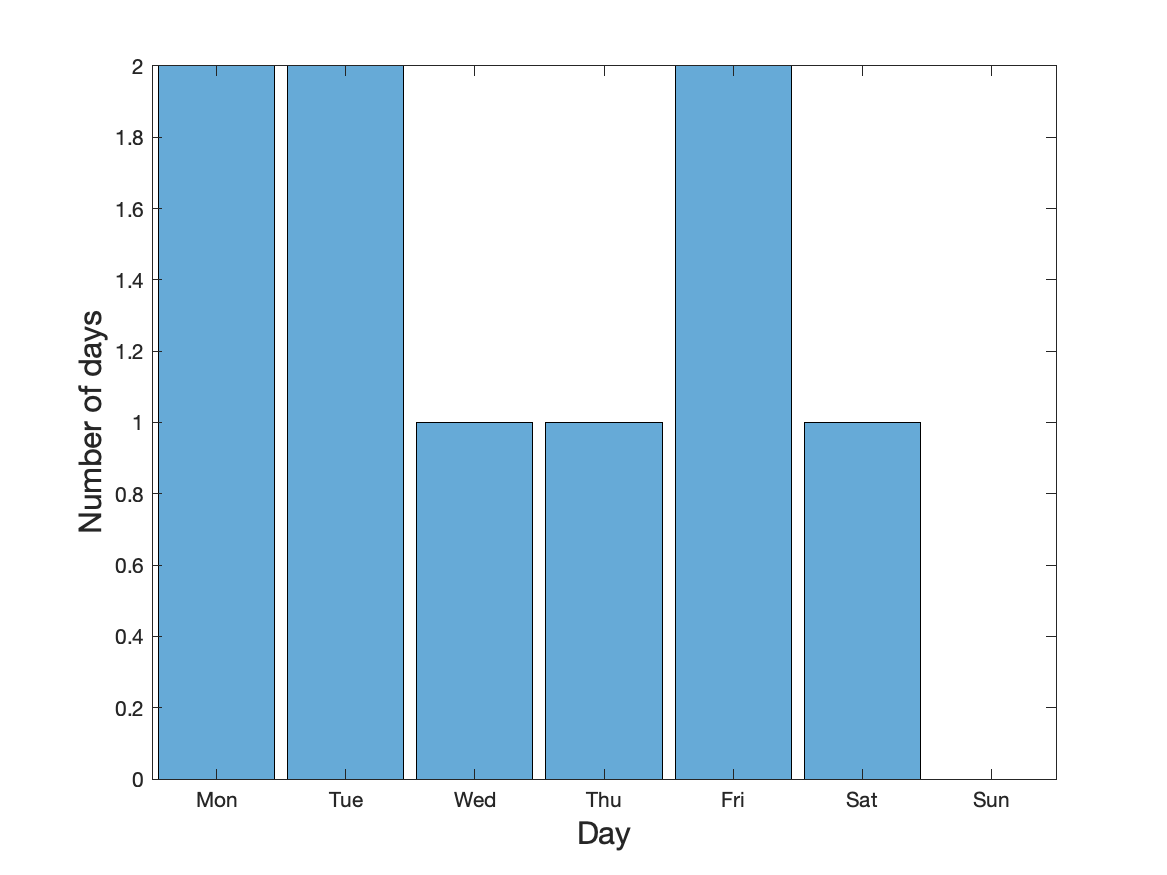}
    \end{minipage} \\
\hline     
2 &
 \begin{minipage}{.3\textwidth}
 \vspace*{0.02in}
  \includegraphics[trim = 20 0 0 0, width=\linewidth, height=29mm]{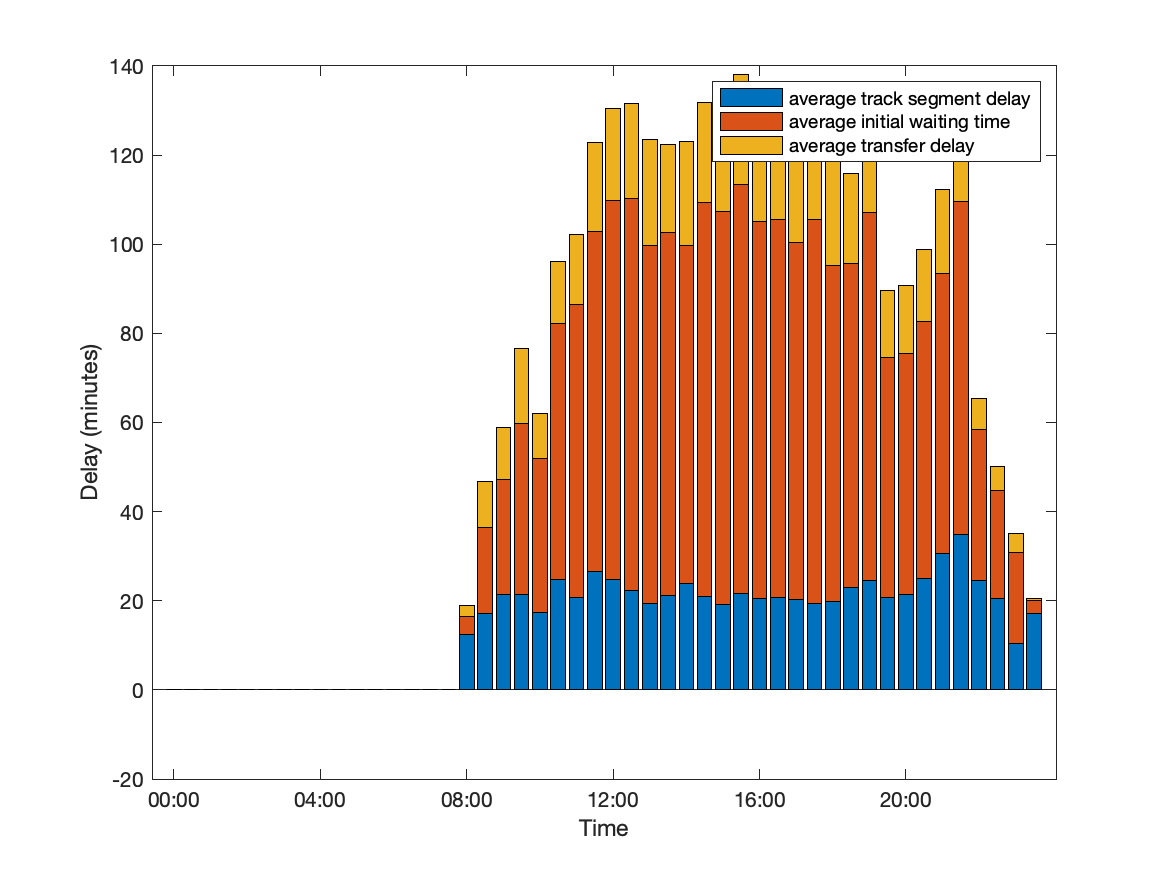}
    \end{minipage} & 
 \begin{minipage}{.25\textwidth}
 \vspace*{0.02in}      \includegraphics[trim = 20 0 0 0, width=\linewidth, height=29mm]{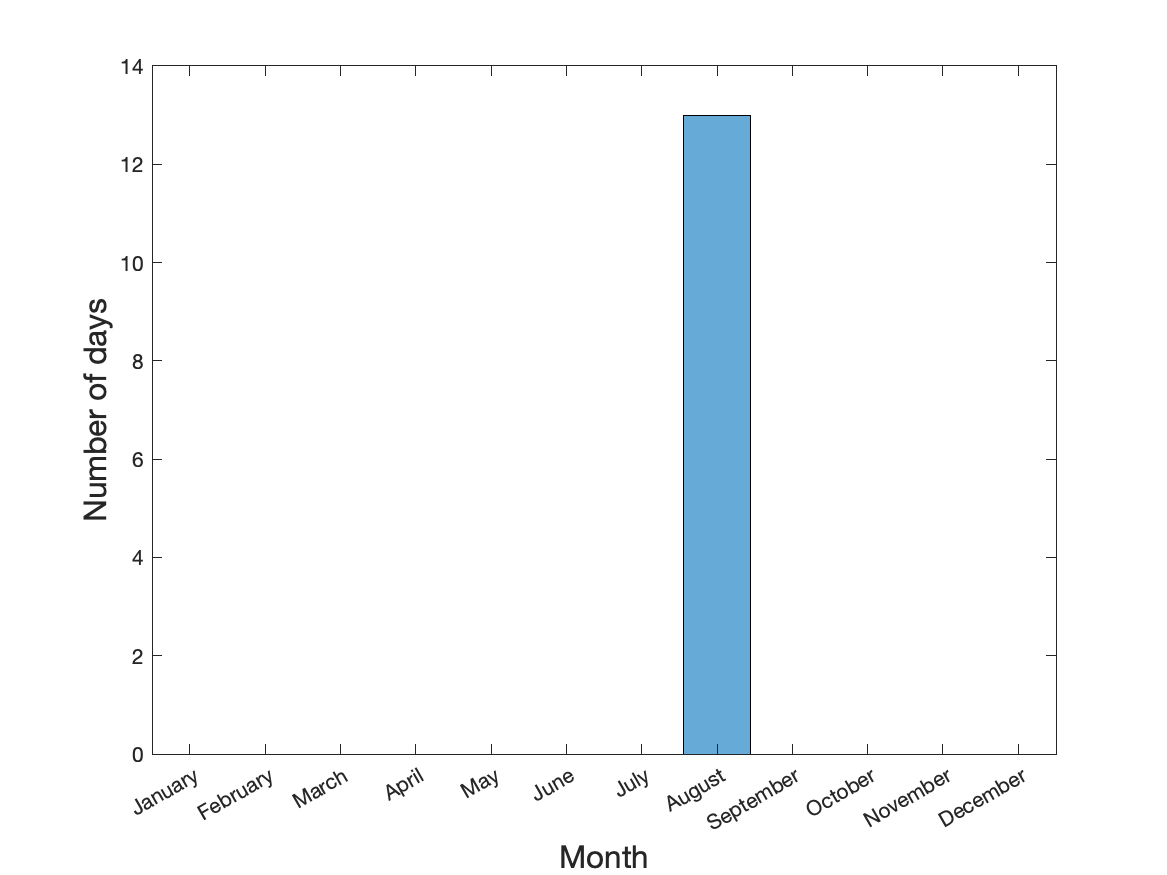}
    \end{minipage}  & 
 \begin{minipage}{.25\textwidth}
 \vspace*{0.02in}   \includegraphics[trim = 20 0 0 0, width=\linewidth, height=29mm]{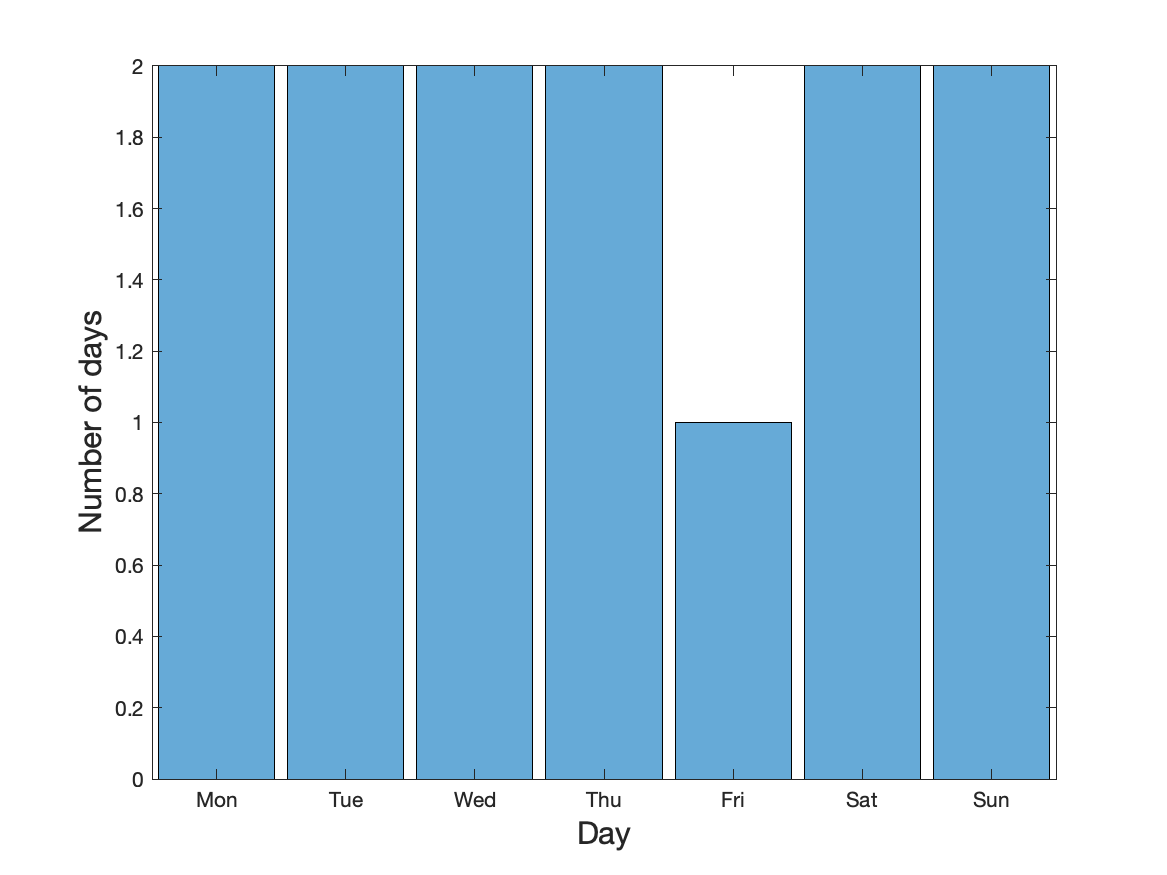}
    \end{minipage} \\
\hline 
3 &
 \begin{minipage}{.3\textwidth}
 \vspace*{0.02in}
  \includegraphics[trim = 20 0 0 0, width=\linewidth, height=29mm]{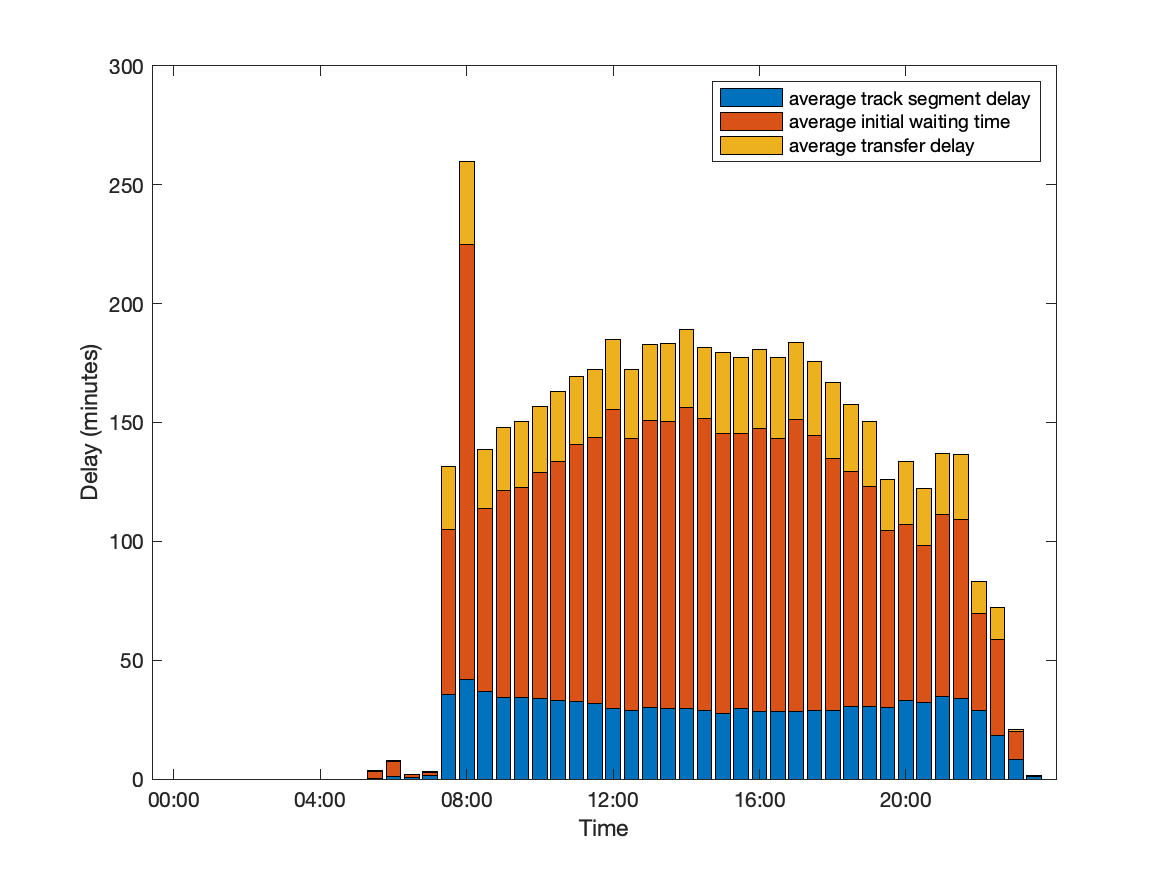}
    \end{minipage} & 
 \begin{minipage}{.25\textwidth}
 \vspace*{0.02in}      \includegraphics[trim = 20 0 0 0, width=\linewidth, height=29mm]{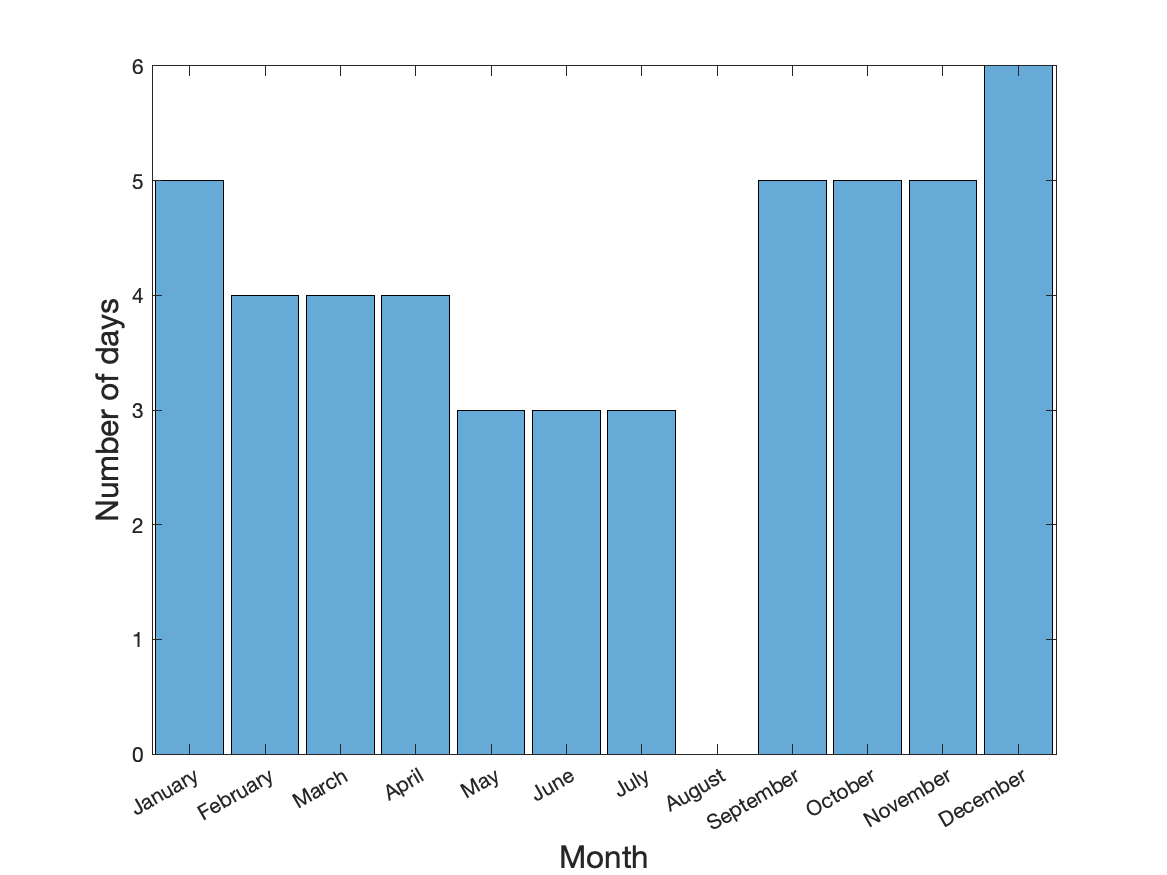}
    \end{minipage}  & 
 \begin{minipage}{.25\textwidth}
 \vspace*{0.02in}   \includegraphics[trim = 20 0 0 0, width=\linewidth, height=29mm]{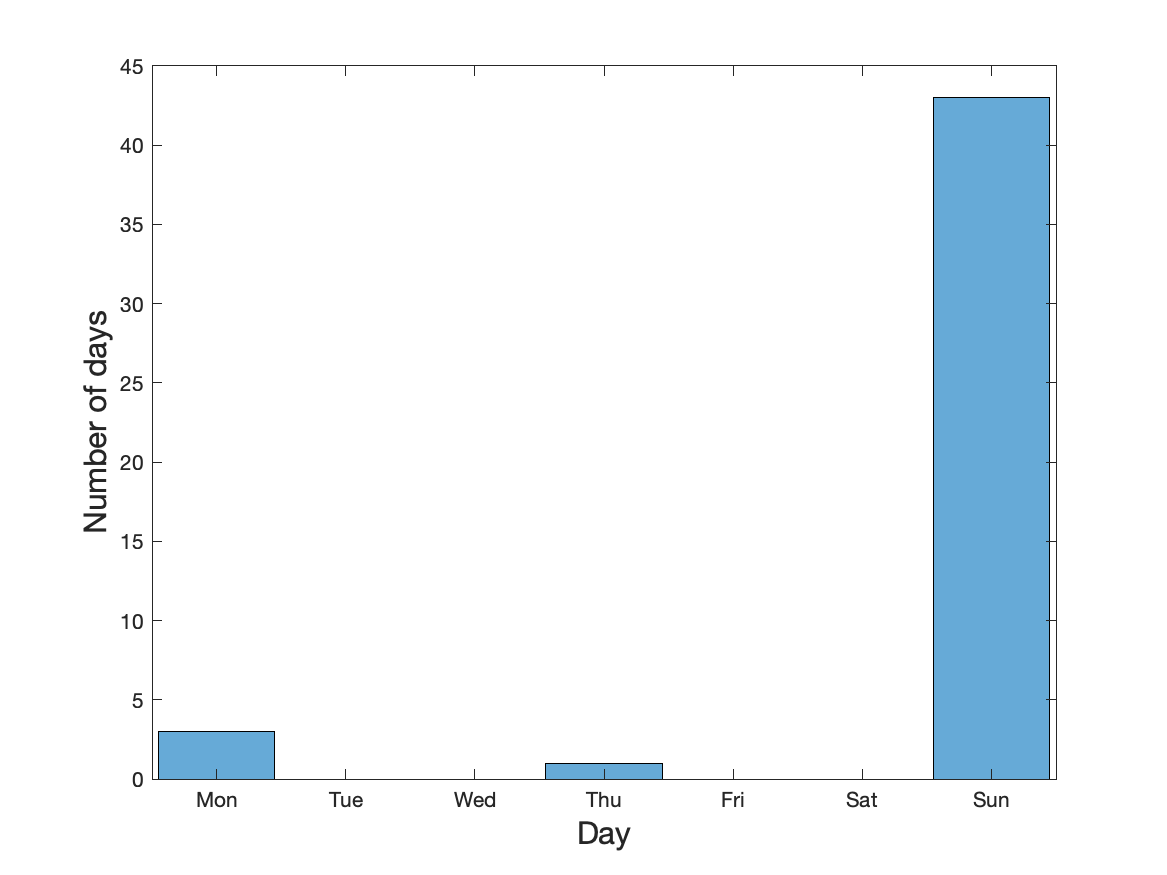}
    \end{minipage} \\
\hline
4 &
 \begin{minipage}{.3\textwidth}
 \vspace*{0.02in}
  \includegraphics[trim = 20 0 0 0, width=\linewidth, height=29mm]{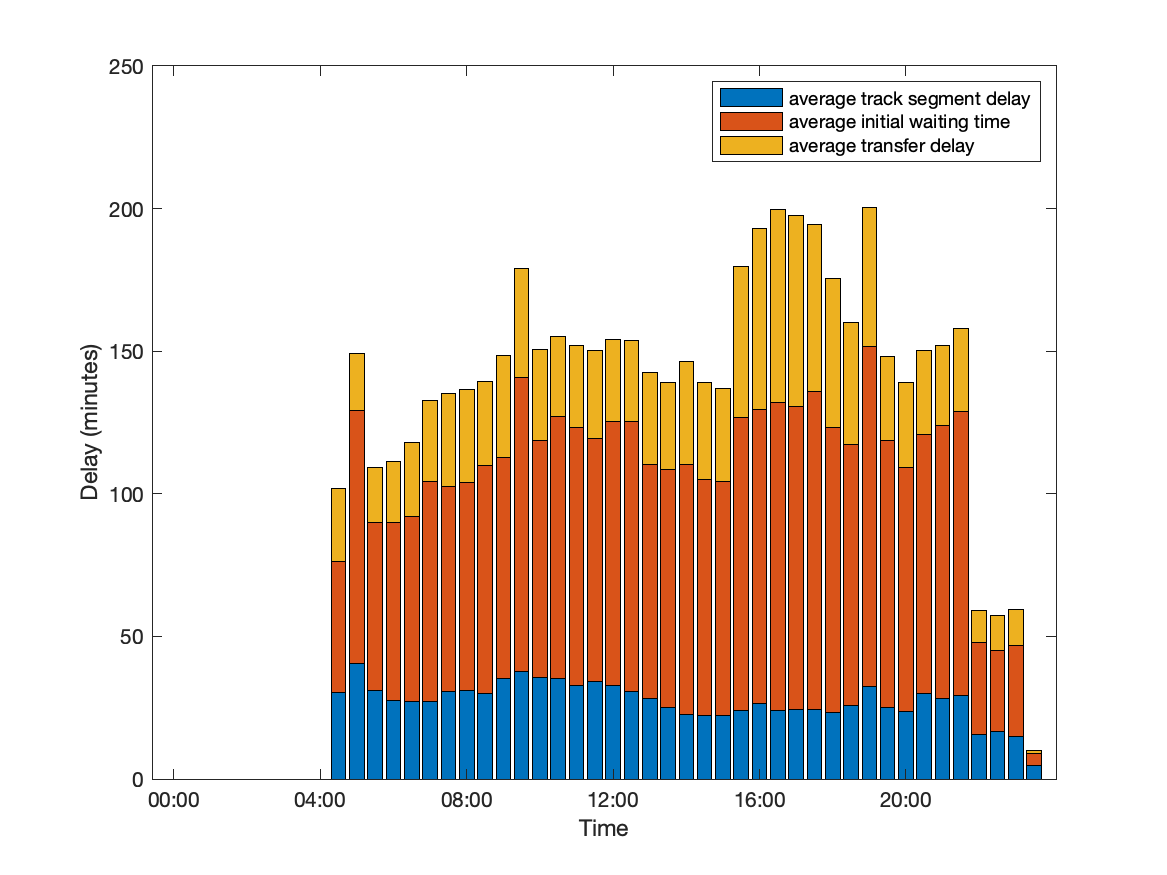}
    \end{minipage} & 
 \begin{minipage}{.25\textwidth}
 \vspace*{0.02in}      \includegraphics[trim = 20 0 0 0, width=\linewidth, height=29mm]{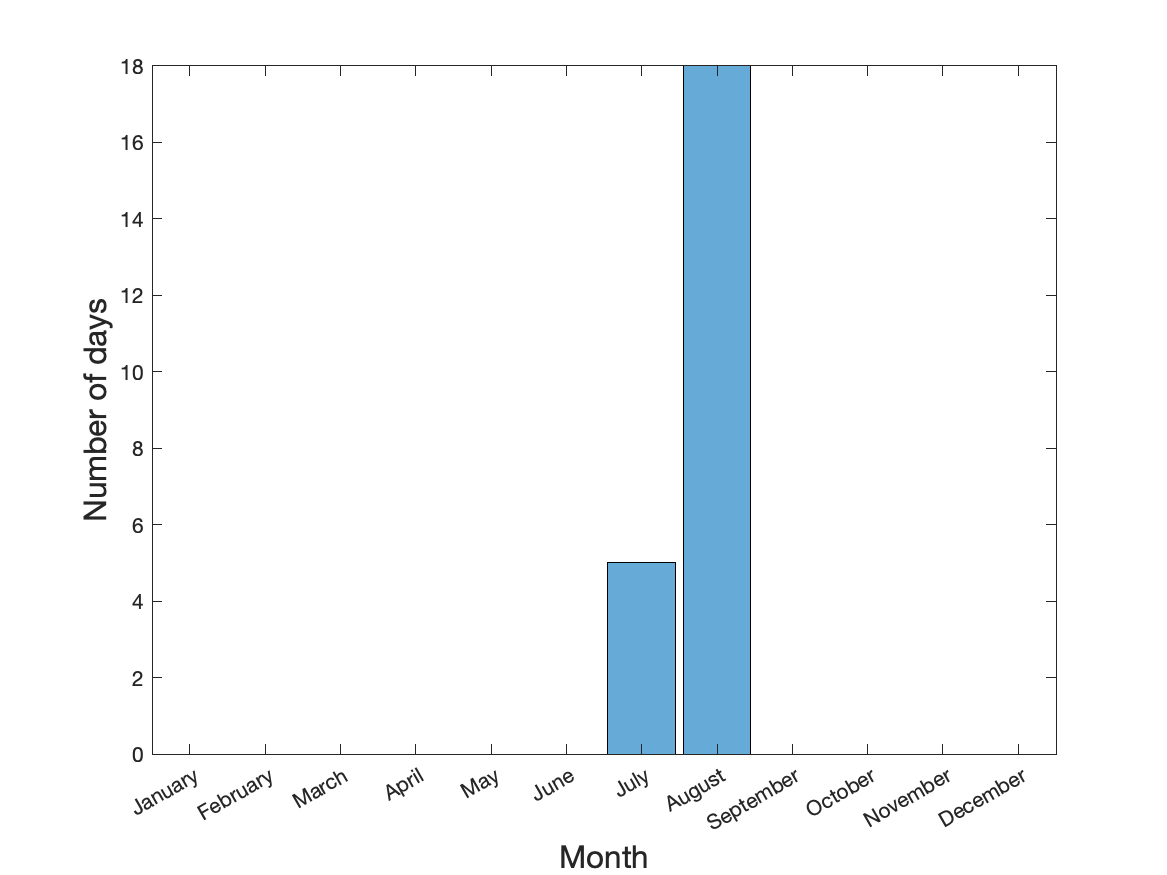}
    \end{minipage}  & 
 \begin{minipage}{.25\textwidth}
 \vspace*{0.02in}   \includegraphics[trim = 20 0 0 0, width=\linewidth, height=29mm]{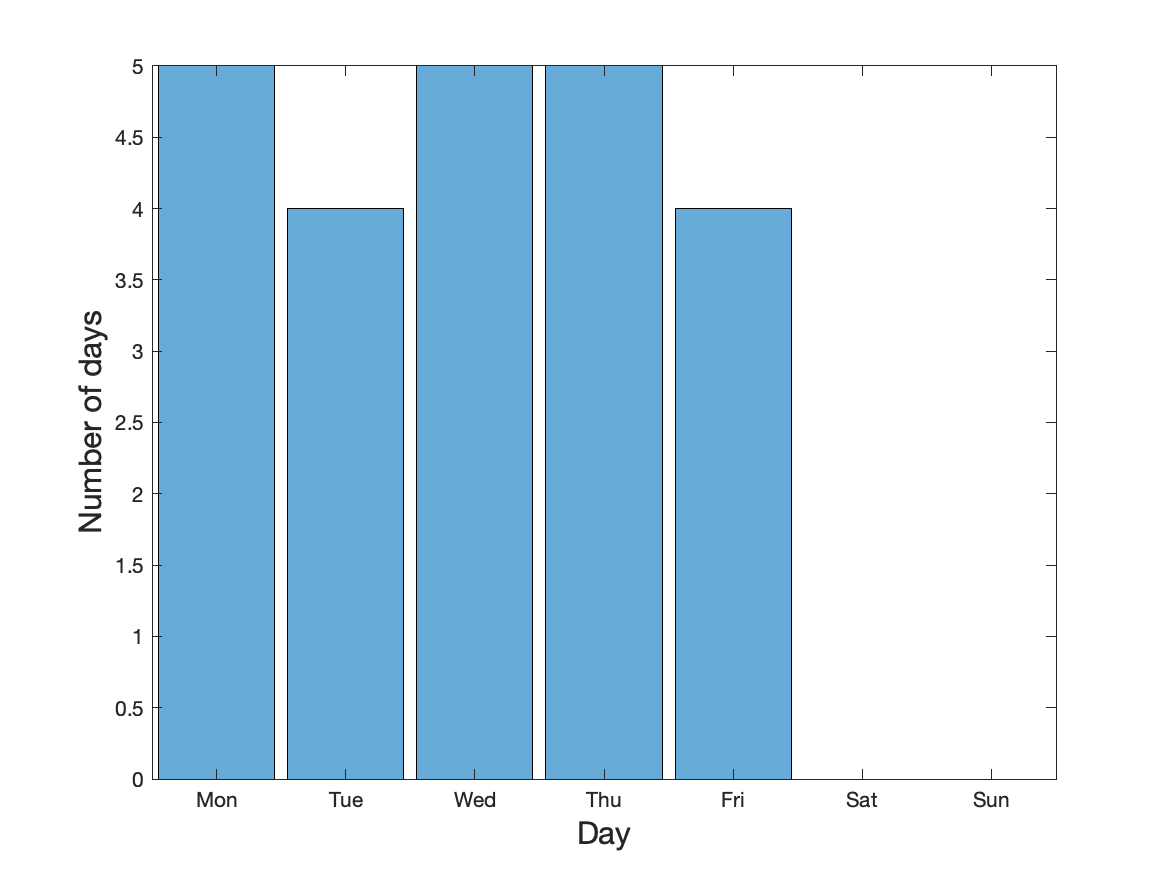}
    \end{minipage} \\
\hline
5 &
 \begin{minipage}{.3\textwidth}
 \vspace*{0.02in}
  \includegraphics[trim = 20 0 0 0, width=\linewidth, height=29mm]{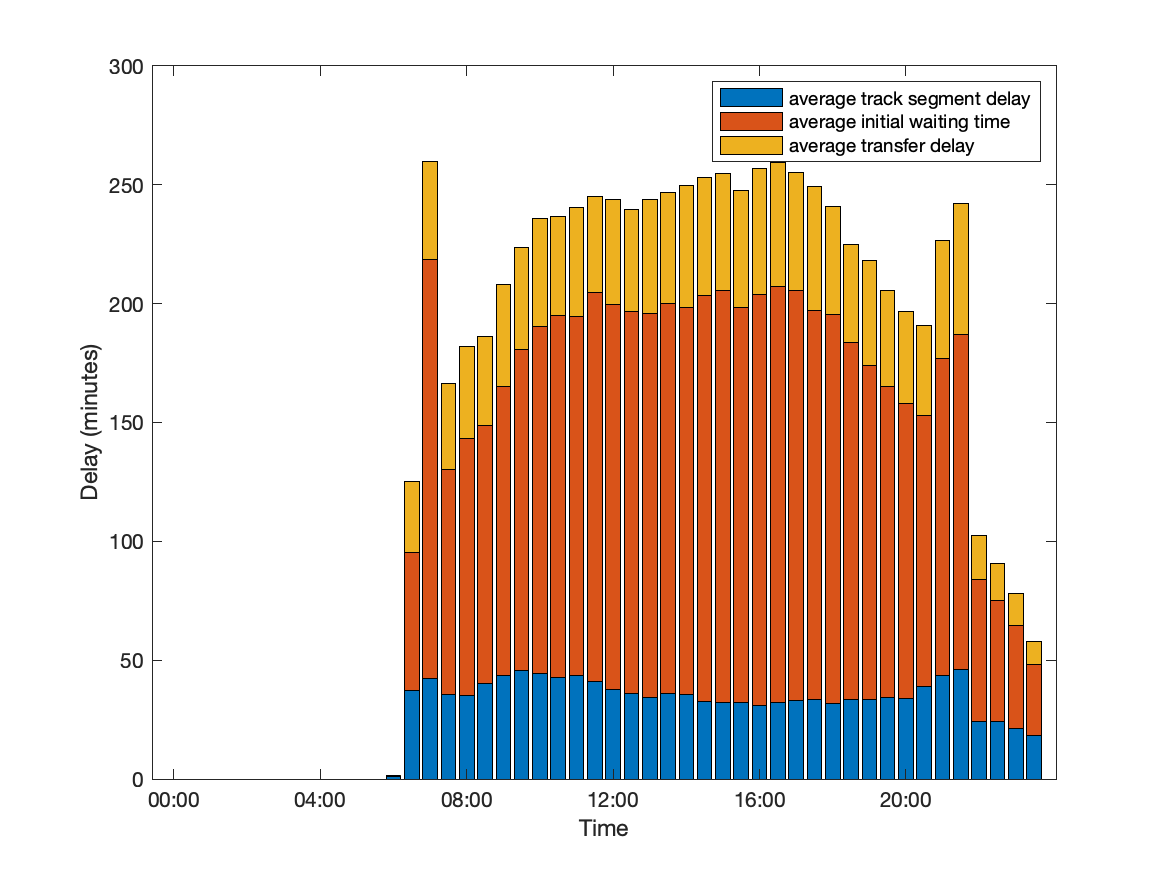}
    \end{minipage} & 
 \begin{minipage}{.25\textwidth}
 \vspace*{0.02in}      \includegraphics[trim = 20 0 0 0, width=\linewidth, height=29mm]{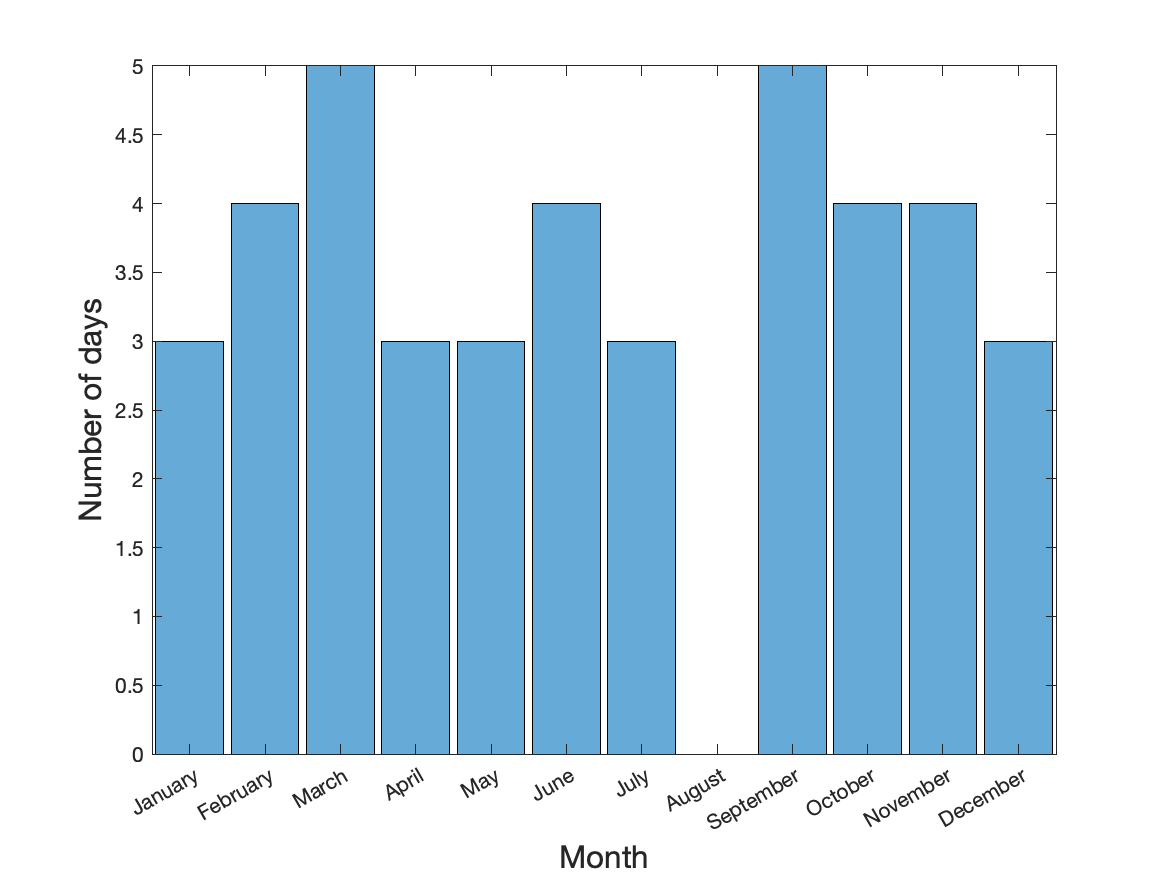}
    \end{minipage}  & 
 \begin{minipage}{.25\textwidth}
 \vspace*{0.02in}   \includegraphics[trim = 20 0 0 0, width=\linewidth, height=29mm]{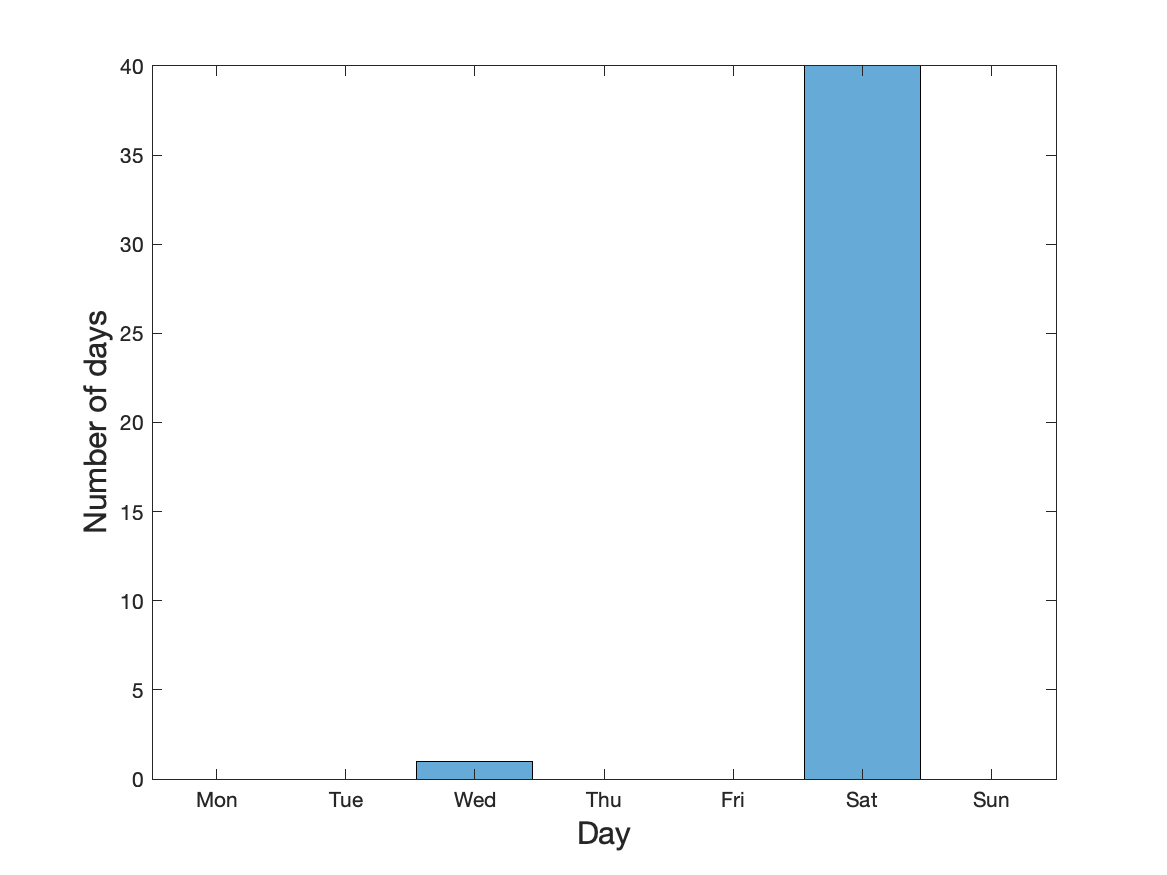}
    \end{minipage} \\
\hline
6 &
 \begin{minipage}{.3\textwidth}
 \vspace*{0.02in}
  \includegraphics[trim = 20 0 0 0, width=\linewidth, height=29mm]{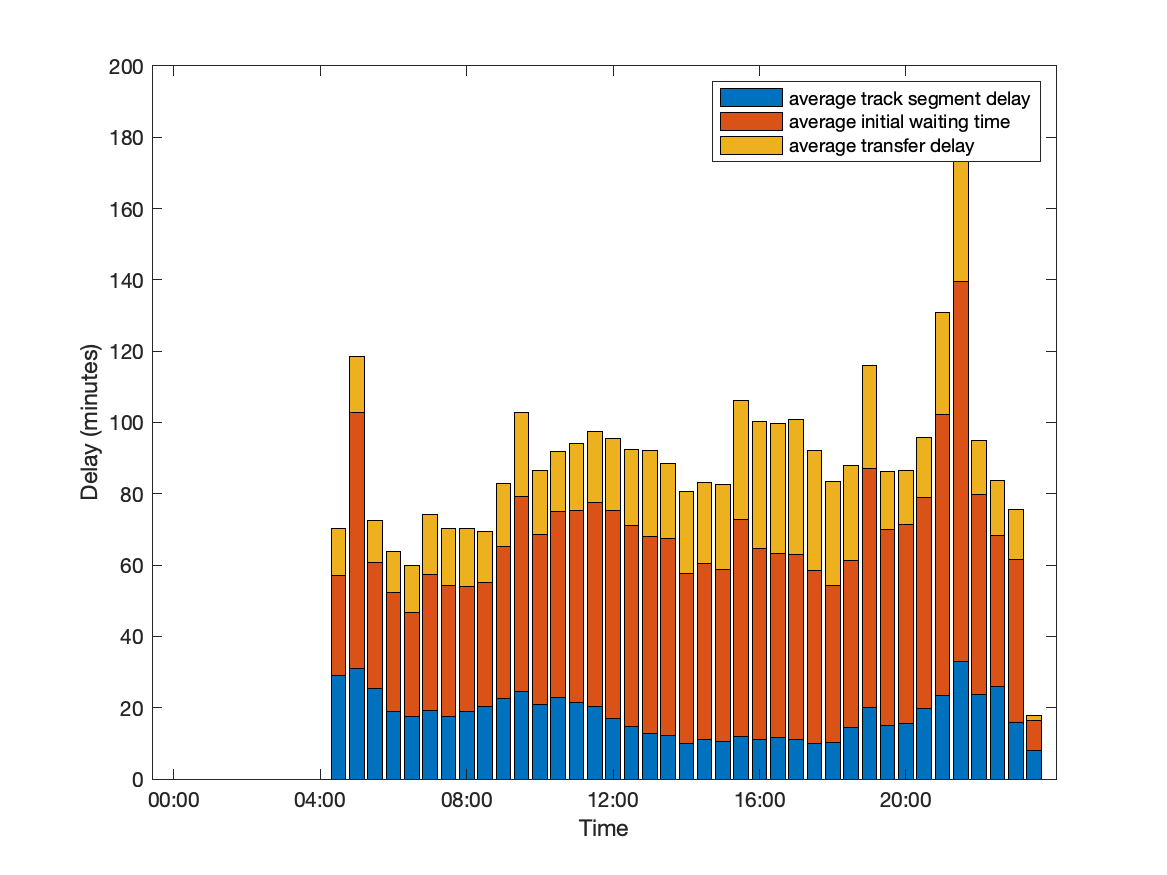}
    \end{minipage} & 
 \begin{minipage}{.25\textwidth}
 \vspace*{0.02in}      \includegraphics[trim = 20 0 0 0, width=\linewidth, height=29mm]{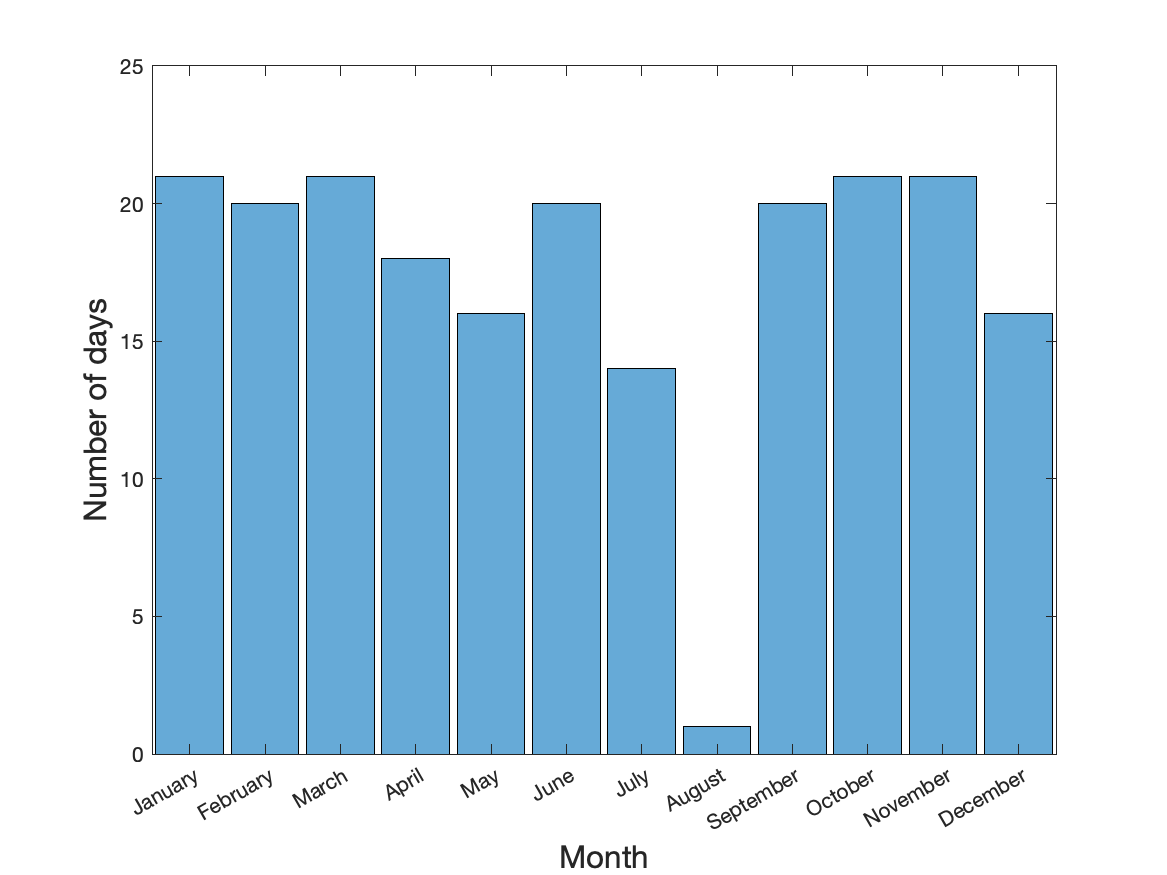}
    \end{minipage}  & 
 \begin{minipage}{.25\textwidth}
 \vspace*{0.02in}   \includegraphics[trim = 20 0 0 0, width=\linewidth, height=29mm]{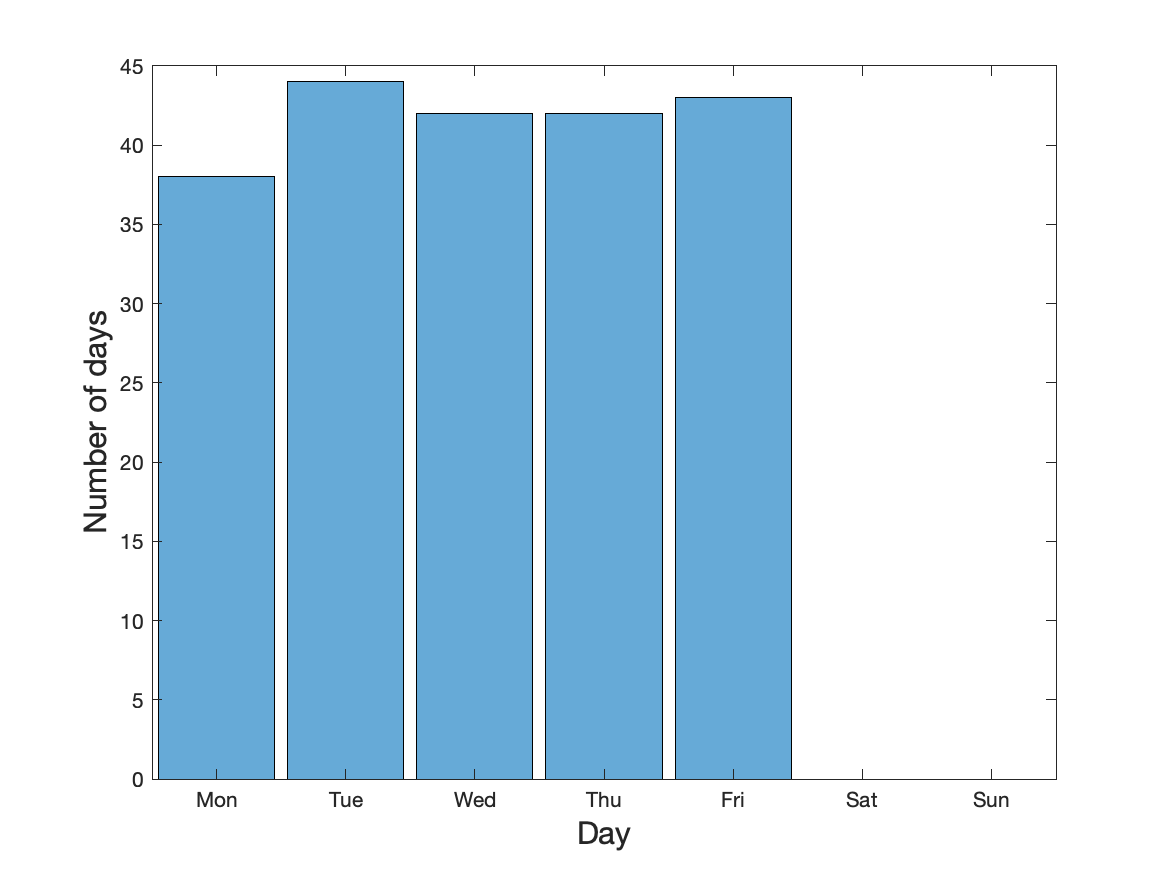}
    \end{minipage} \\
    \hline
7 &
 \begin{minipage}{.3\textwidth}
 \vspace*{0.02in}
  \includegraphics[trim = 20 0 0 0, width=\linewidth, height=29mm]{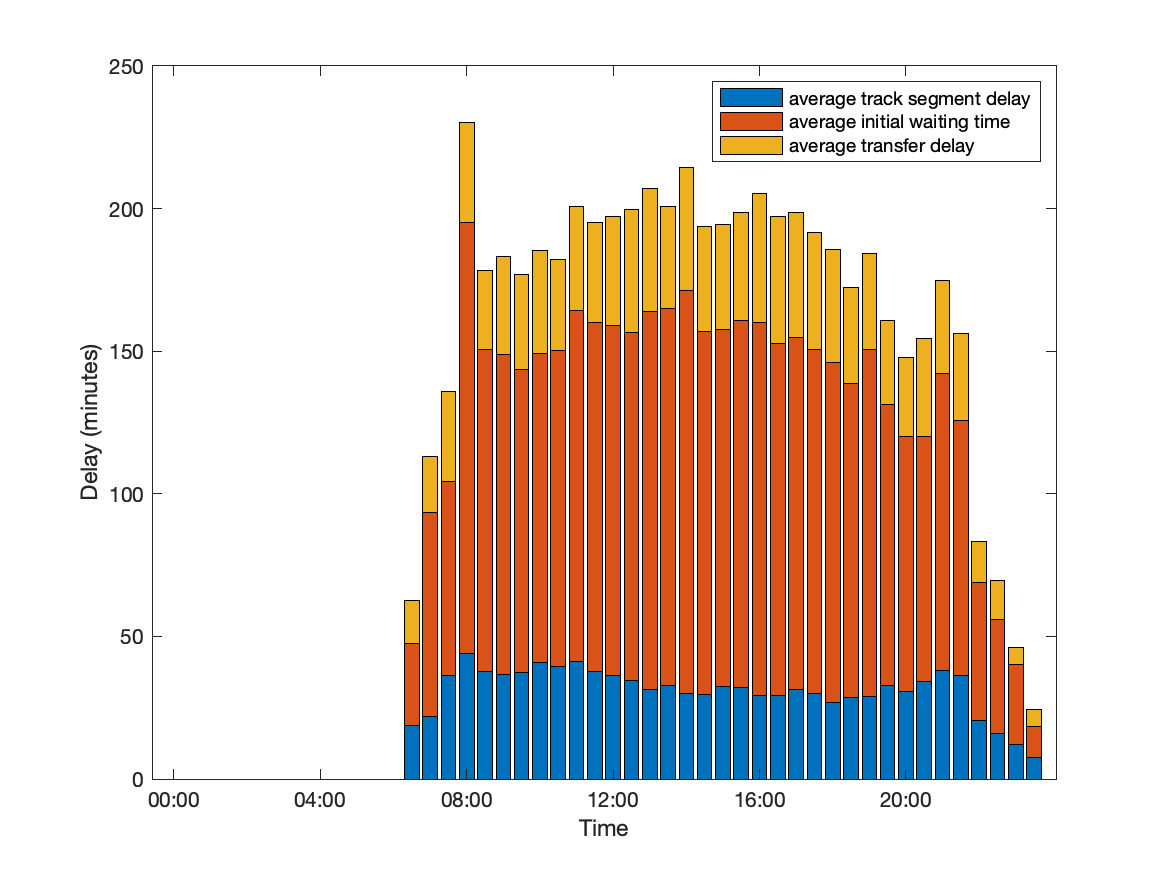}
    \end{minipage} & 
 \begin{minipage}{.25\textwidth}
 \vspace*{0.02in}      \includegraphics[trim = 20 0 0 0, width=\linewidth, height=29mm]{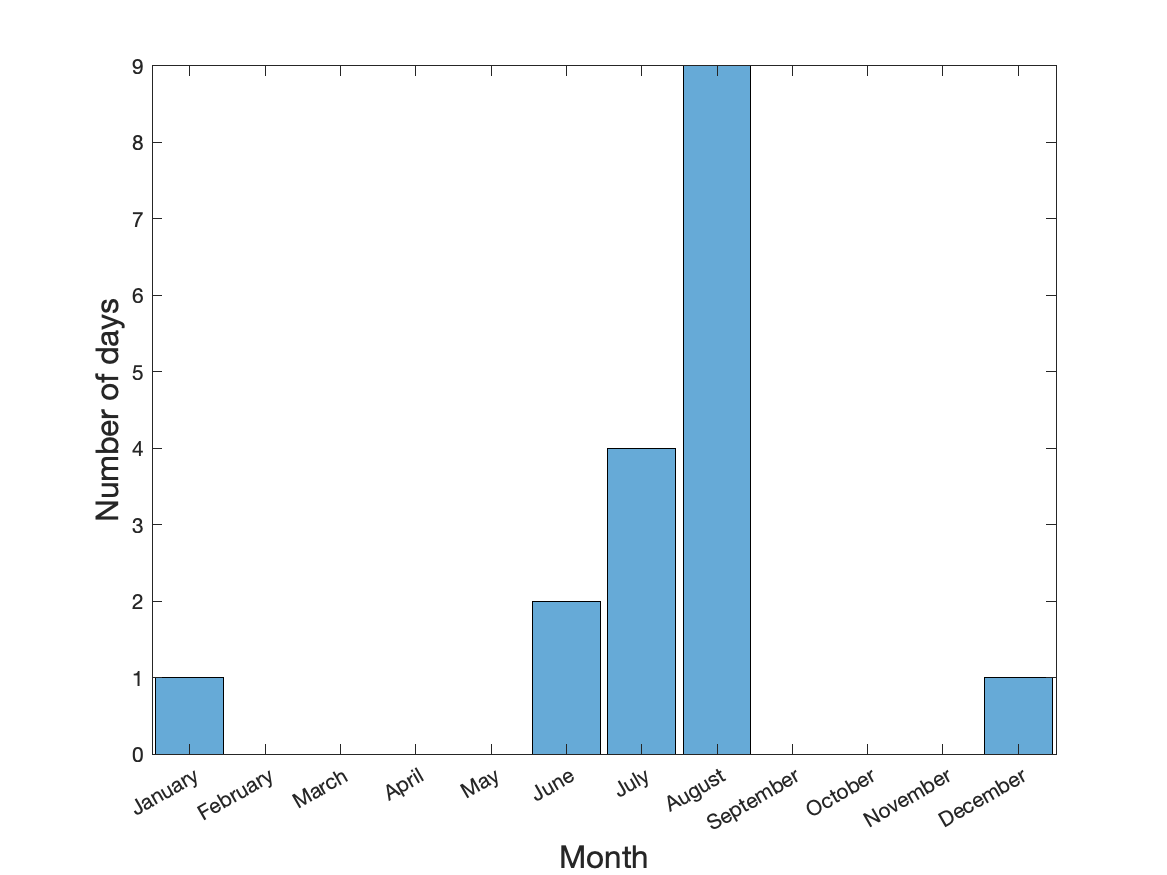}
    \end{minipage}  & 
 \begin{minipage}{.25\textwidth}
 \vspace*{0.02in}   \includegraphics[trim = 20 0 0 0, width=\linewidth, height=29mm]{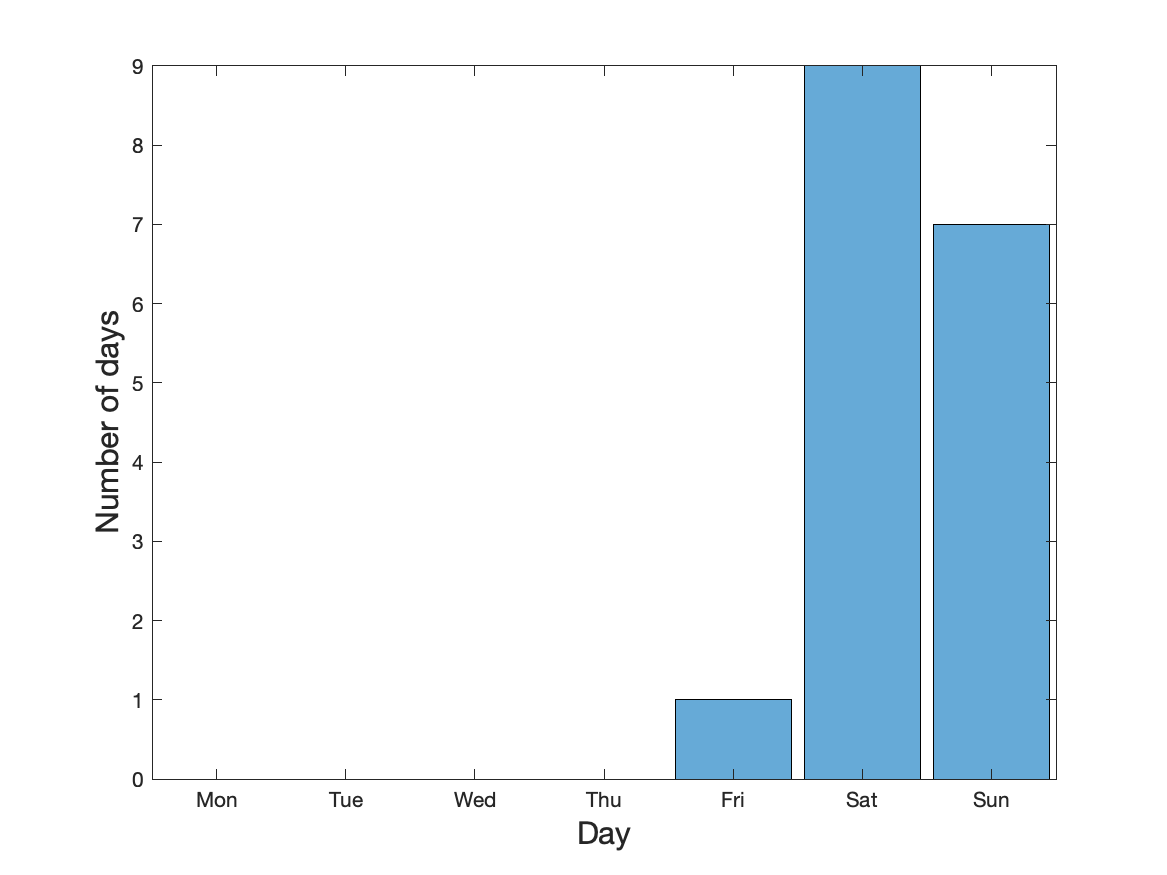}
    \end{minipage} \\
\hline
\end{tabular}
\end{table}

\begin{table}[h!]
\centering
\caption{Description of network states based on total passenger delay.}
\label{classes2}
\begin{tabular} {|c|c|c|c|}
\hline
\textbf{Class} &
\textbf{Delay distribution}  & 
\textbf{Distribution of months} & 
\textbf{Distribution of days}\\ 
\hline
1 &
 \begin{minipage}{.3\textwidth}
 \vspace*{0.02in}
  \includegraphics[trim = 20 0 0 0, width=\linewidth, height=29mm]{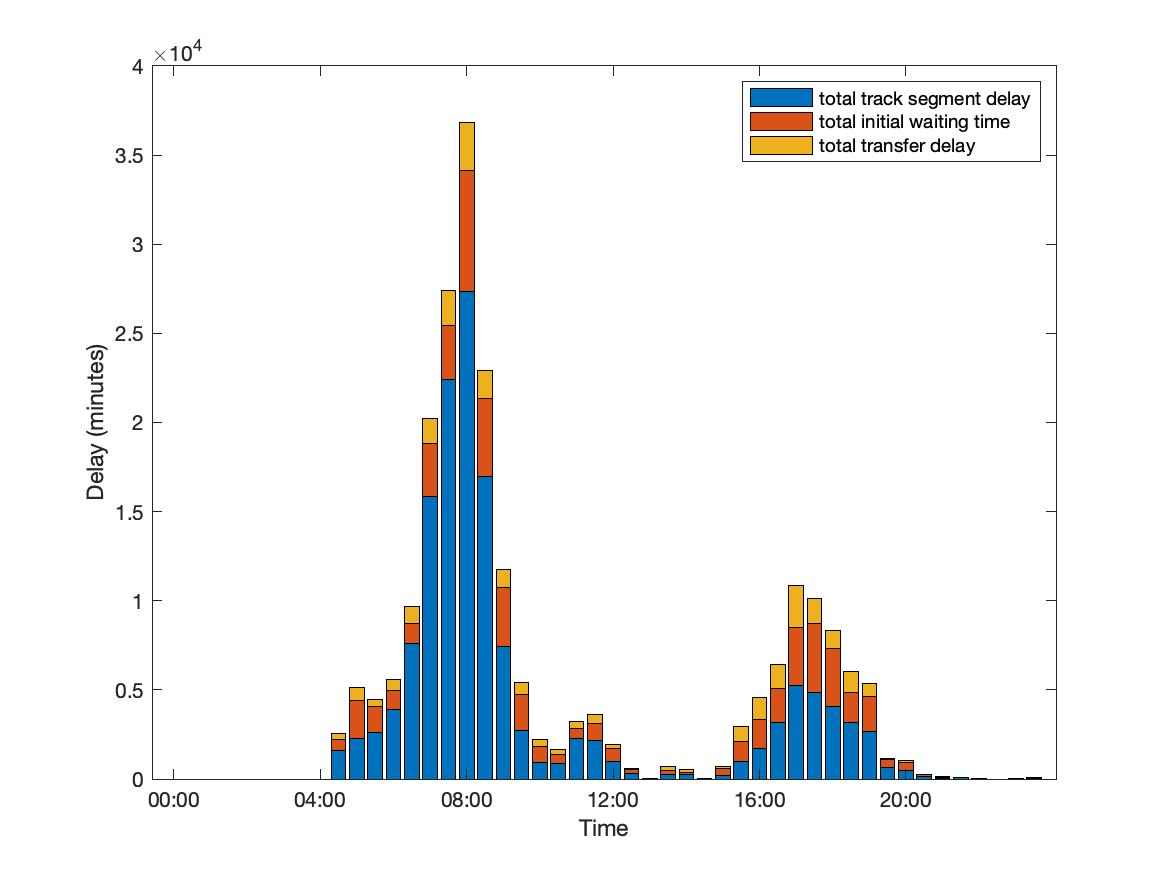}
    \end{minipage} & 
 \begin{minipage}{.25\textwidth}
 \vspace*{0.02in}      \includegraphics[trim = 20 0 0 0, width=\linewidth, height=29mm]{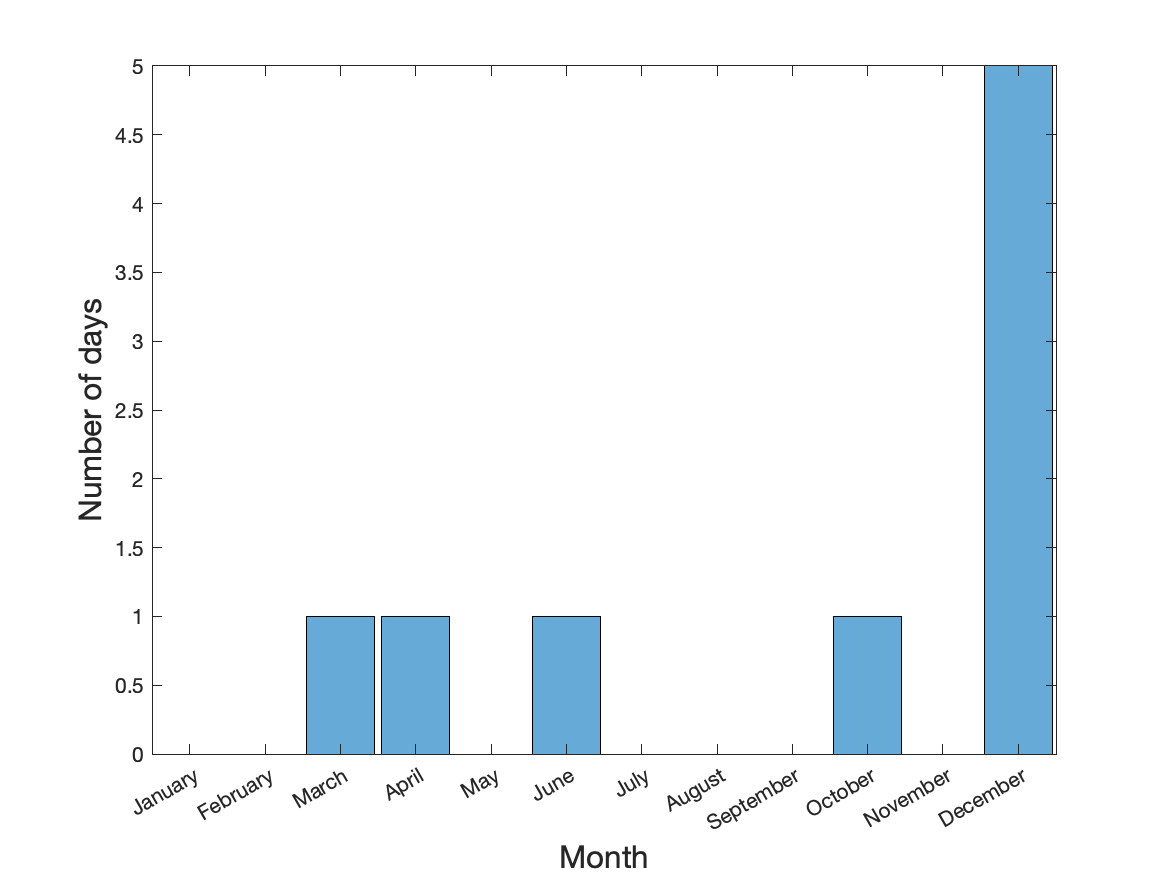}
    \end{minipage}  & 
 \begin{minipage}{.25\textwidth}
 \vspace*{0.02in}   \includegraphics[trim = 20 0 0 0, width=\linewidth, height=29mm]{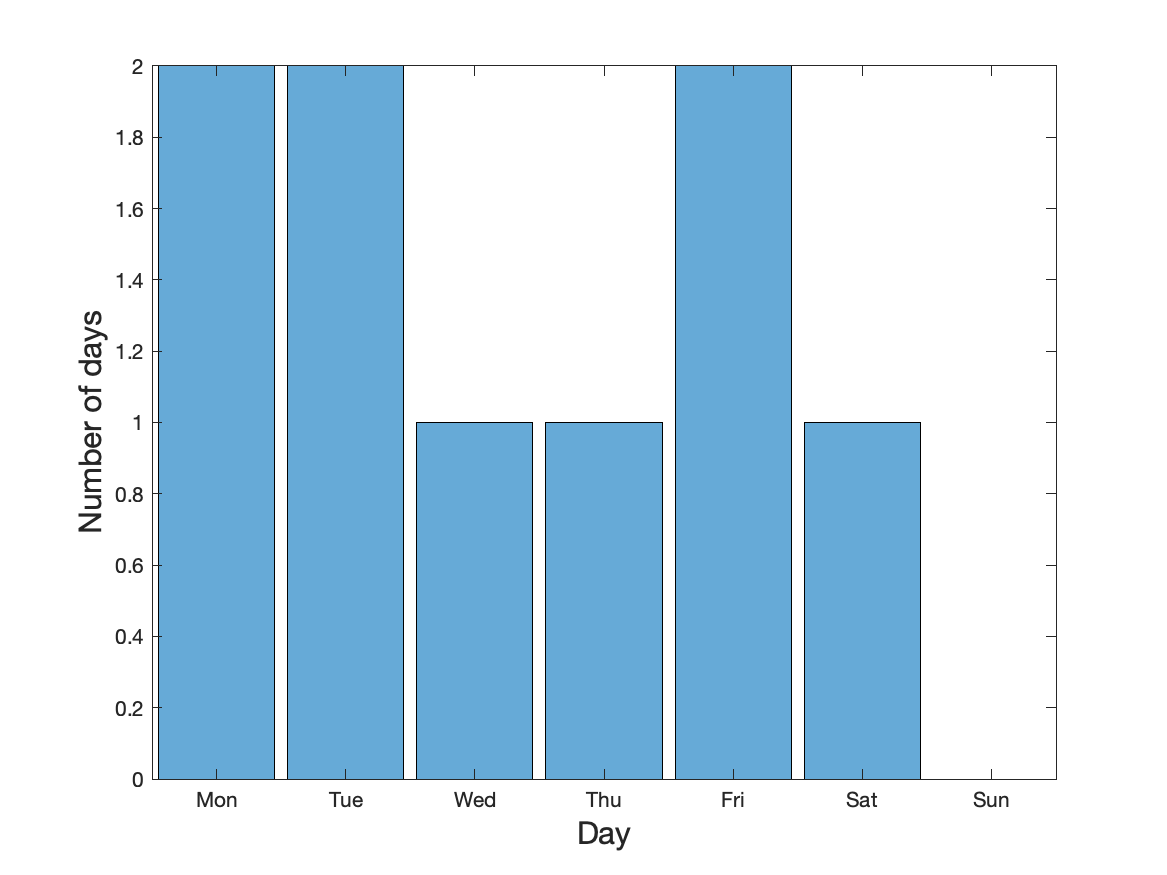}
    \end{minipage} \\
\hline
2 &
 \begin{minipage}{.3\textwidth}
 \vspace*{0.02in}
  \includegraphics[trim = 20 0 0 0, width=\linewidth, height=29mm]{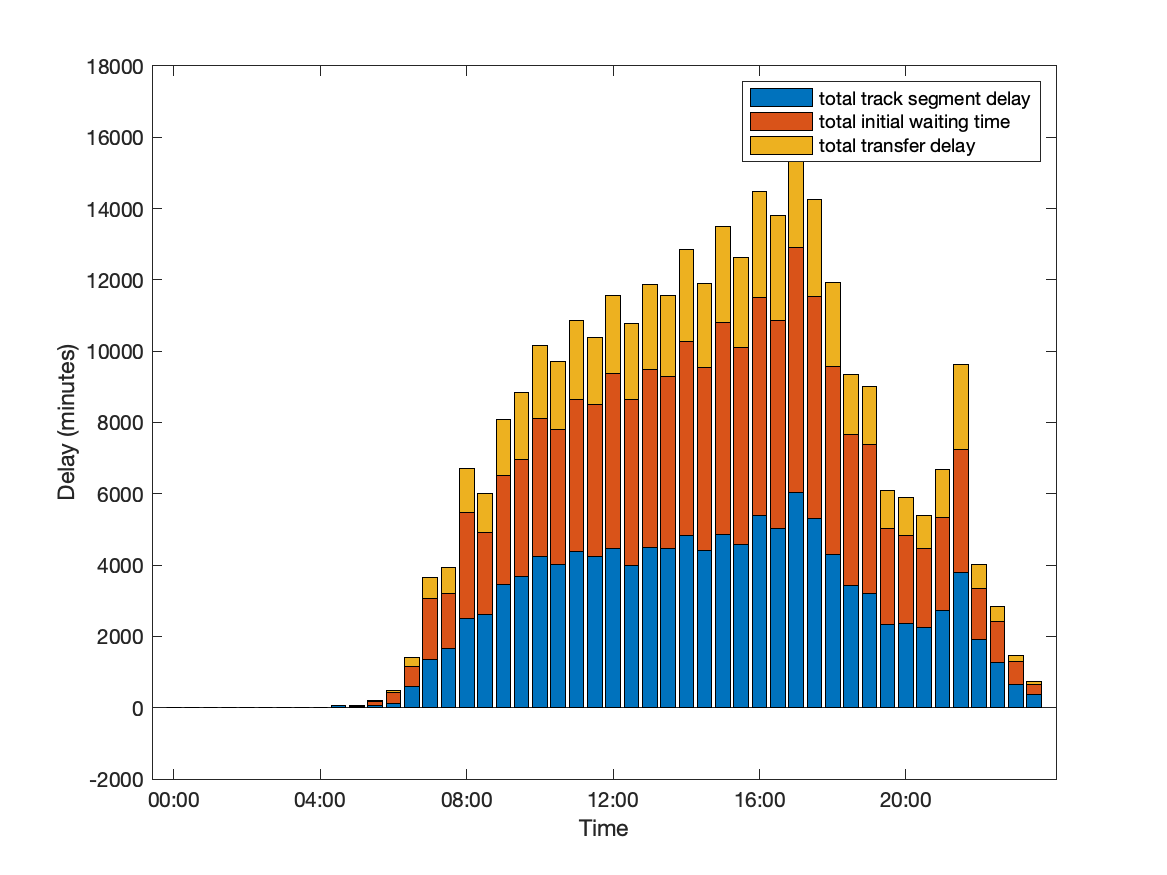}
    \end{minipage} & 
 \begin{minipage}{.25\textwidth}
 \vspace*{0.02in}      \includegraphics[trim = 20 0 0 0, width=\linewidth, height=29mm]{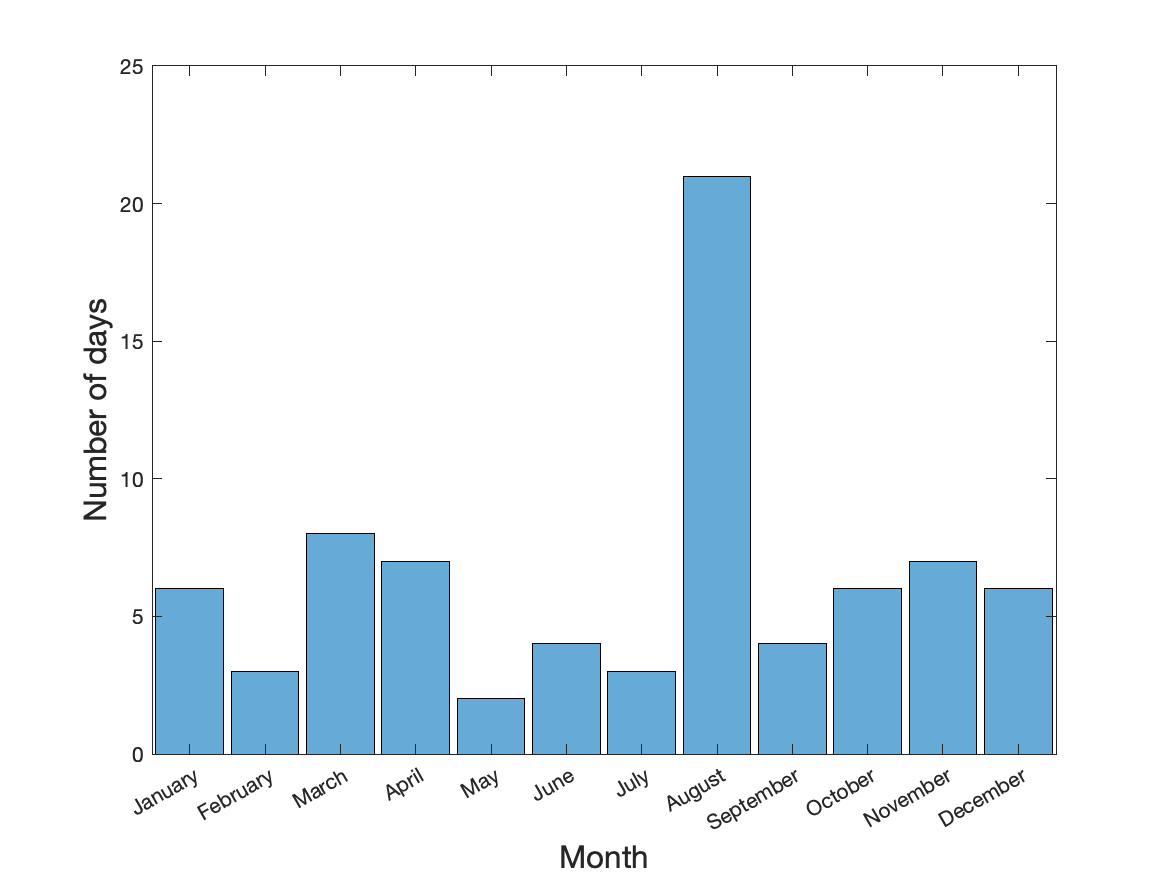}
    \end{minipage}  & 
 \begin{minipage}{.25\textwidth}
 \vspace*{0.02in}   \includegraphics[trim = 20 0 0 0, width=\linewidth, height=29mm]{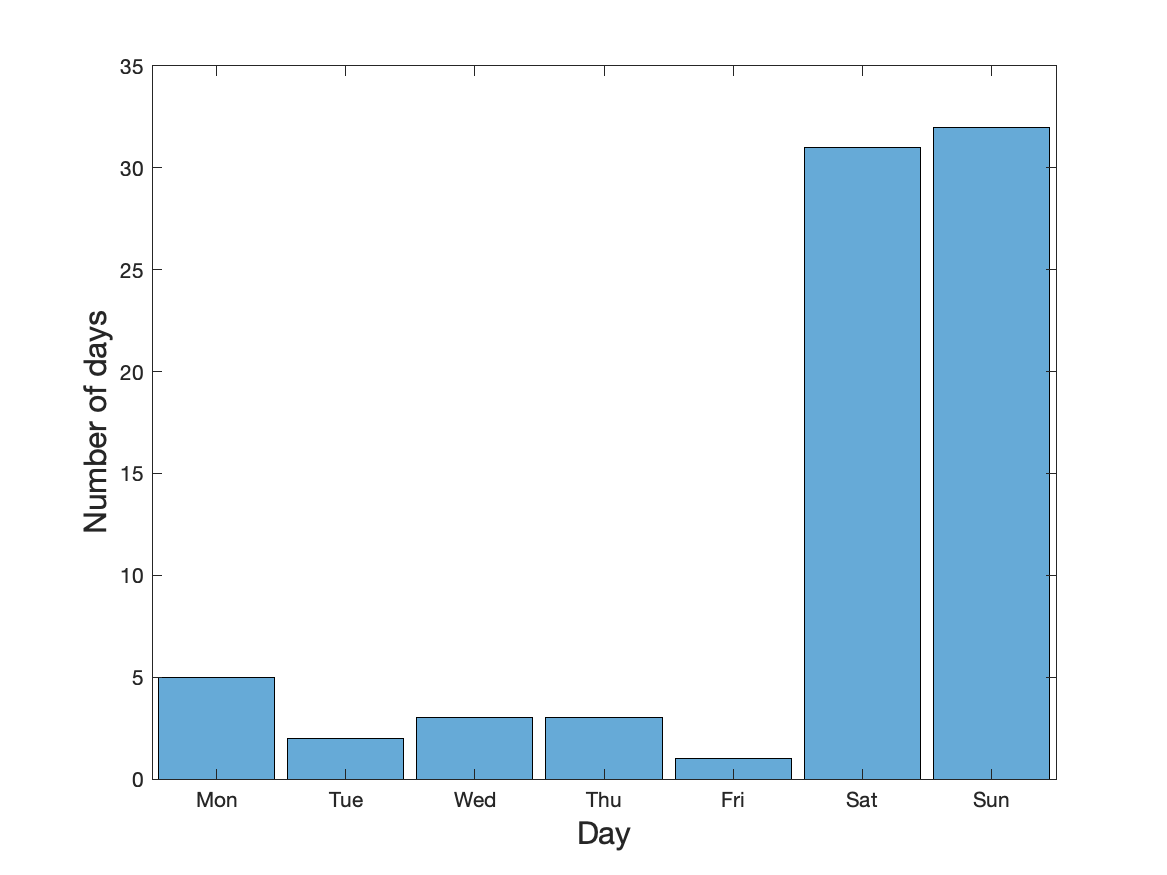}
    \end{minipage} \\
\hline 
3 &
 \begin{minipage}{.3\textwidth}
 \vspace*{0.02in}
  \includegraphics[trim = 20 0 0 0, width=\linewidth, height=29mm]{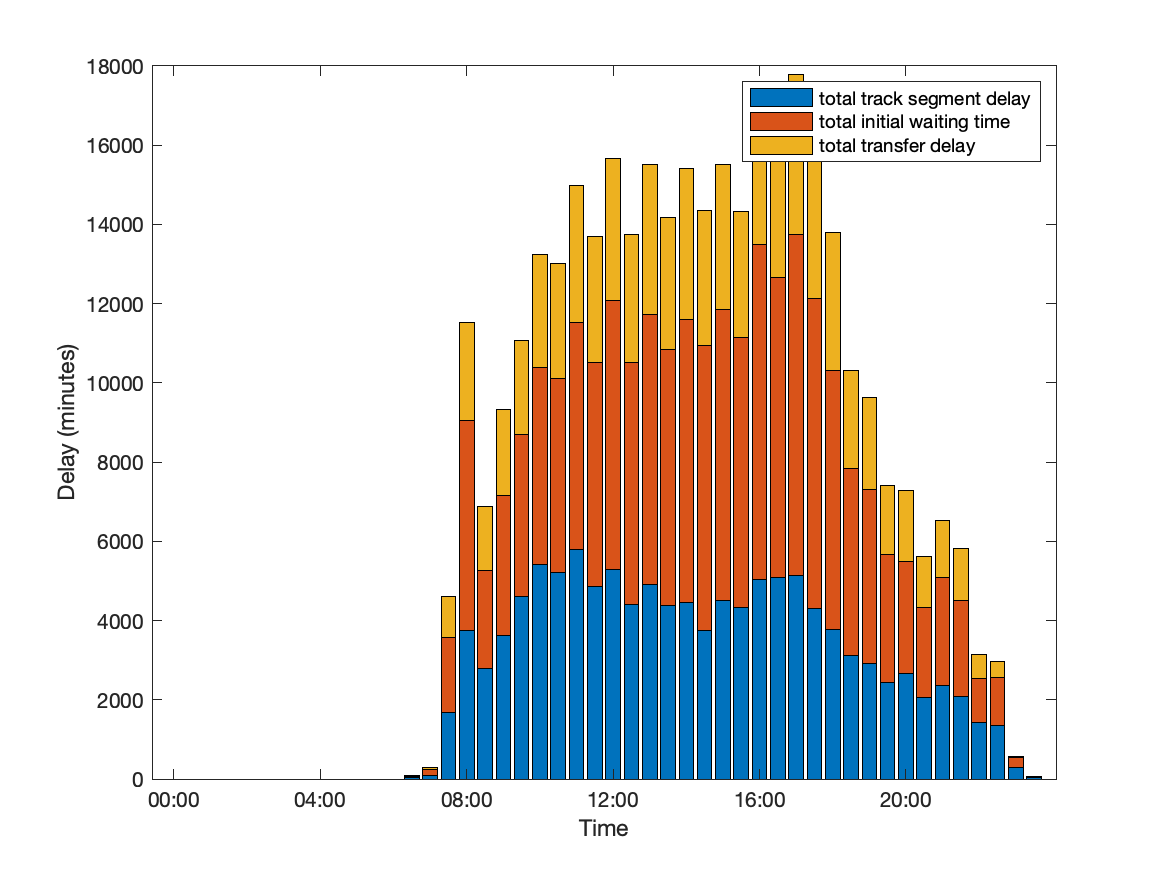}
    \end{minipage} & 
 \begin{minipage}{.25\textwidth}
 \vspace*{0.02in}      \includegraphics[trim = 20 0 0 0, width=\linewidth, height=29mm]{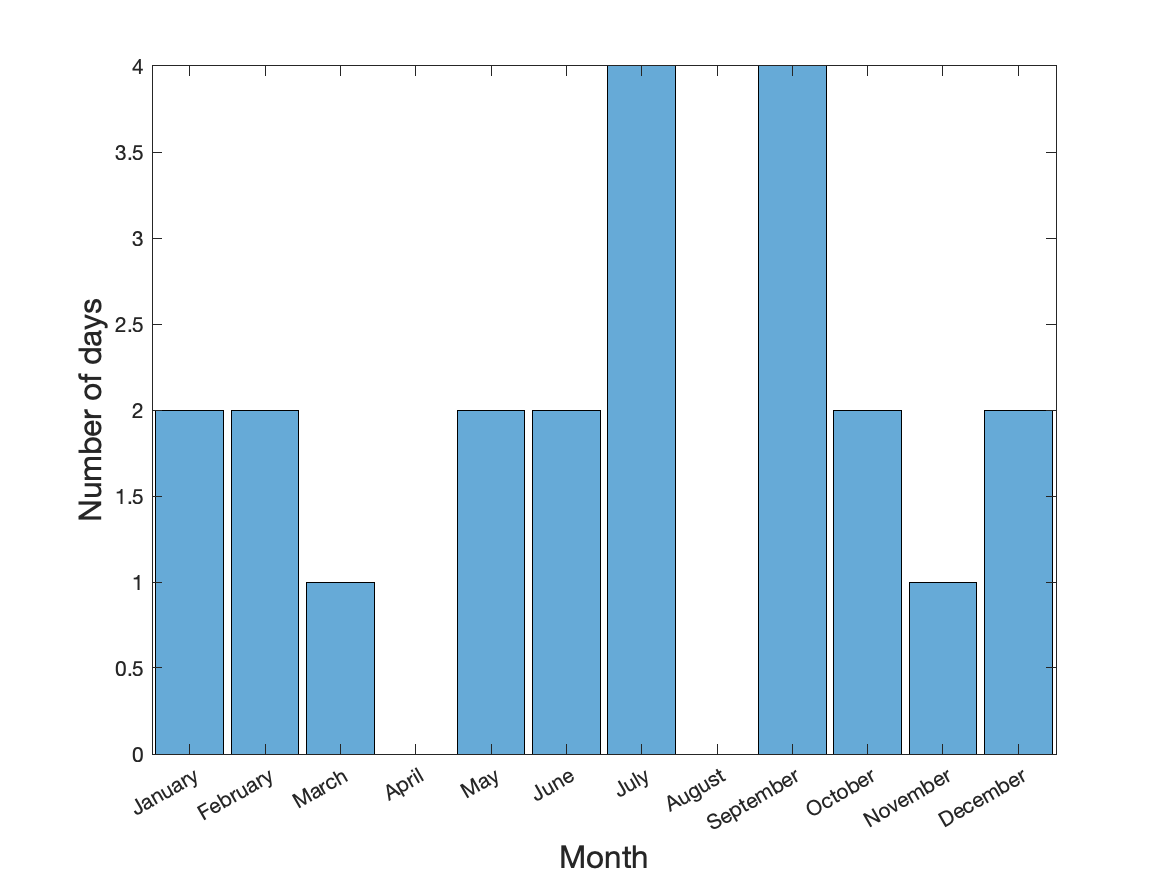}
    \end{minipage}  & 
 \begin{minipage}{.25\textwidth}
 \vspace*{0.02in}   \includegraphics[trim = 20 0 0 0, width=\linewidth, height=29mm]{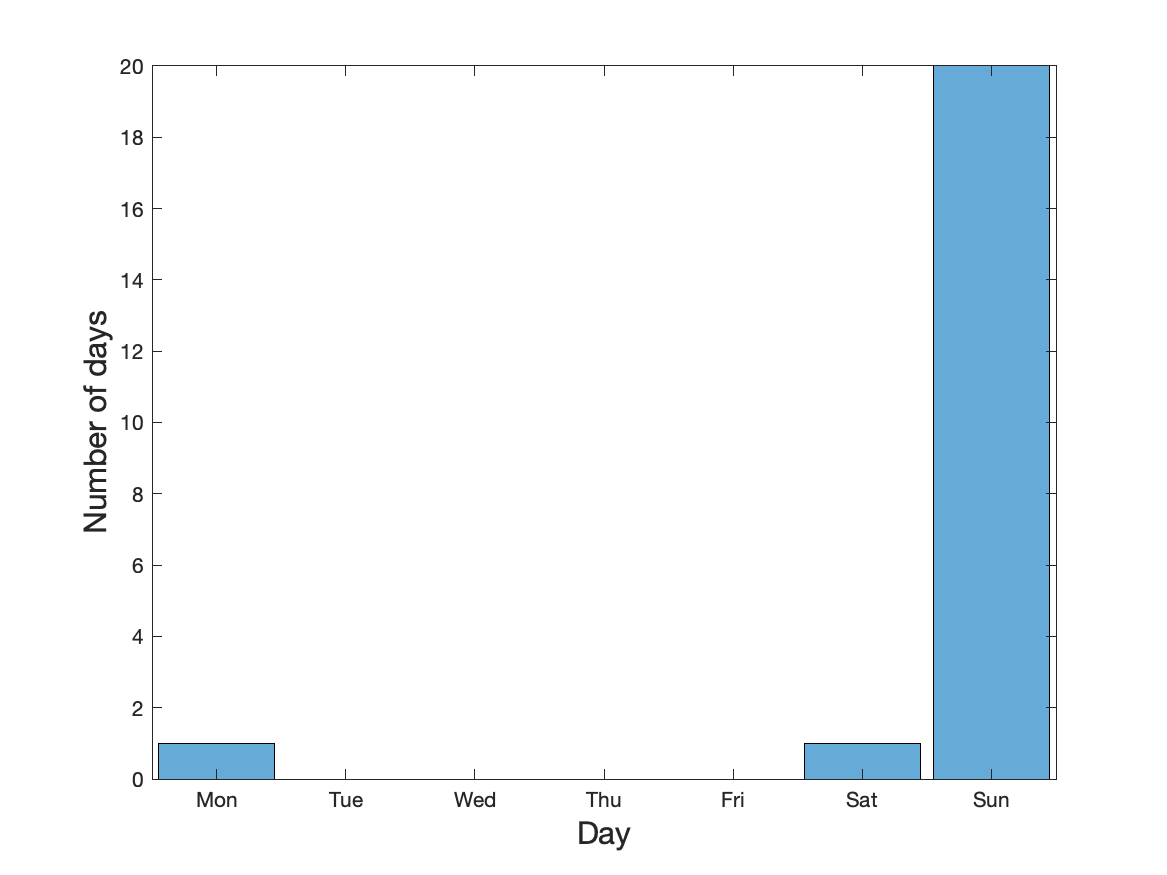}
    \end{minipage} \\
\hline
4 &
 \begin{minipage}{.3\textwidth}
 \vspace*{0.02in}
  \includegraphics[trim = 20 0 0 0, width=\linewidth, height=29mm]{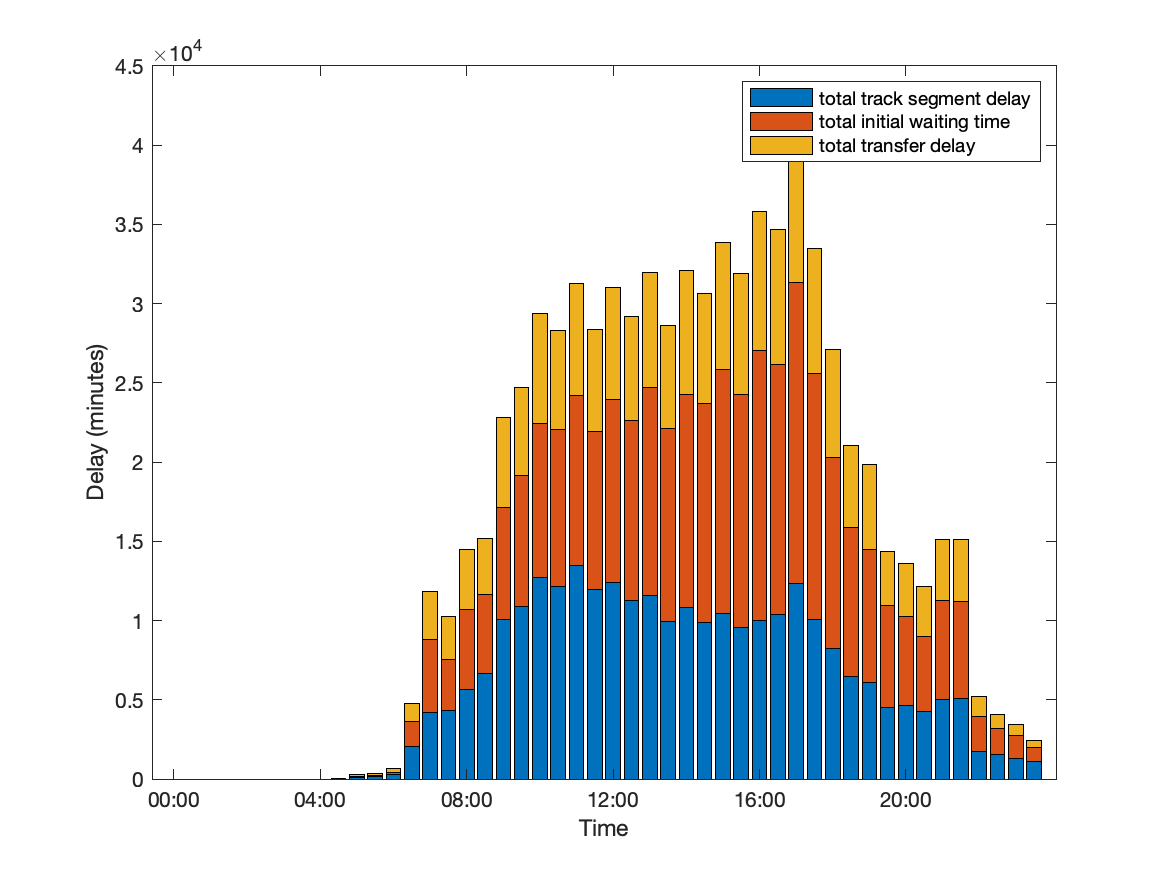}
    \end{minipage} & 
 \begin{minipage}{.25\textwidth}
 \vspace*{0.02in}      \includegraphics[trim = 20 0 0 0, width=\linewidth, height=29mm]{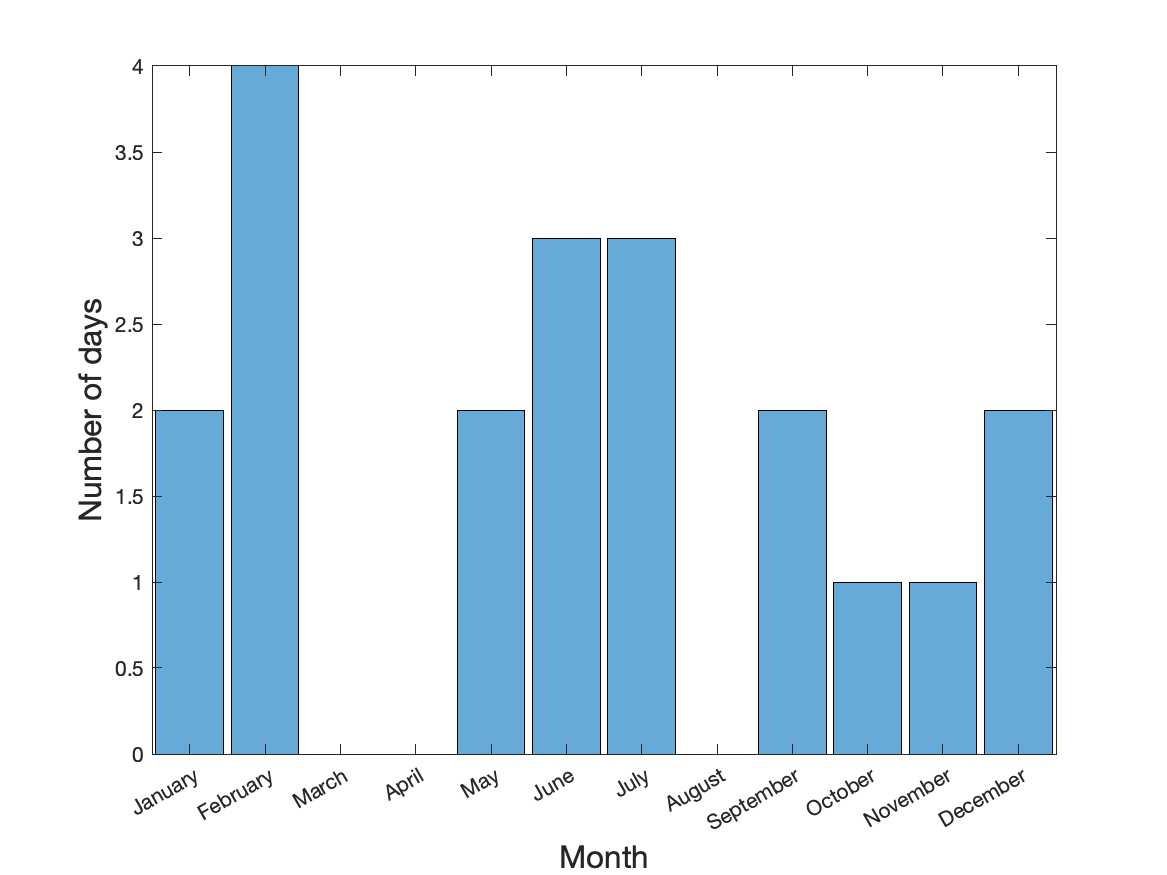}
    \end{minipage}  & 
 \begin{minipage}{.25\textwidth}
 \vspace*{0.02in}   \includegraphics[trim = 20 0 0 0, width=\linewidth, height=29mm]{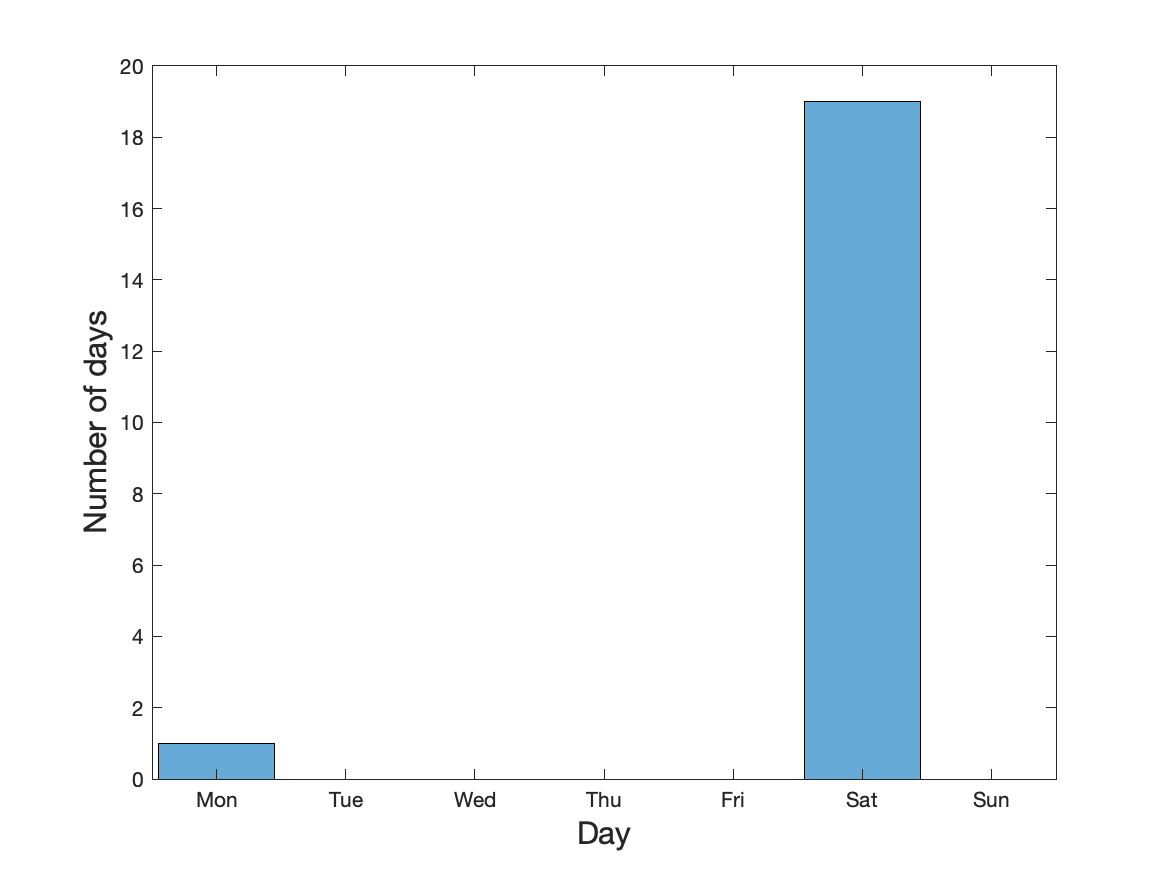}
    \end{minipage} \\
\hline
5 &
 \begin{minipage}{.3\textwidth}
 \vspace*{0.02in}
  \includegraphics[trim = 20 0 0 0, width=\linewidth, height=29mm]{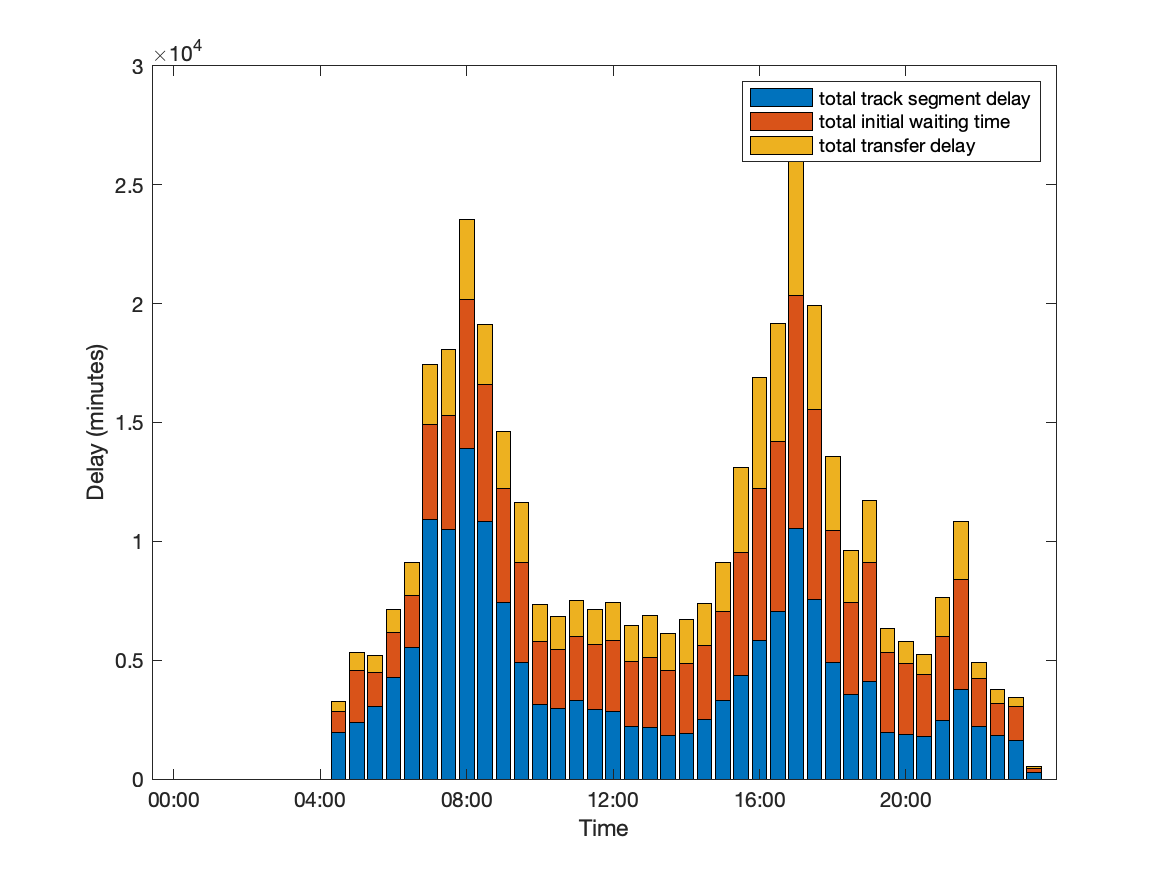}
    \end{minipage} & 
 \begin{minipage}{.25\textwidth}
 \vspace*{0.02in}      \includegraphics[trim = 20 0 0 0, width=\linewidth, height=29mm]{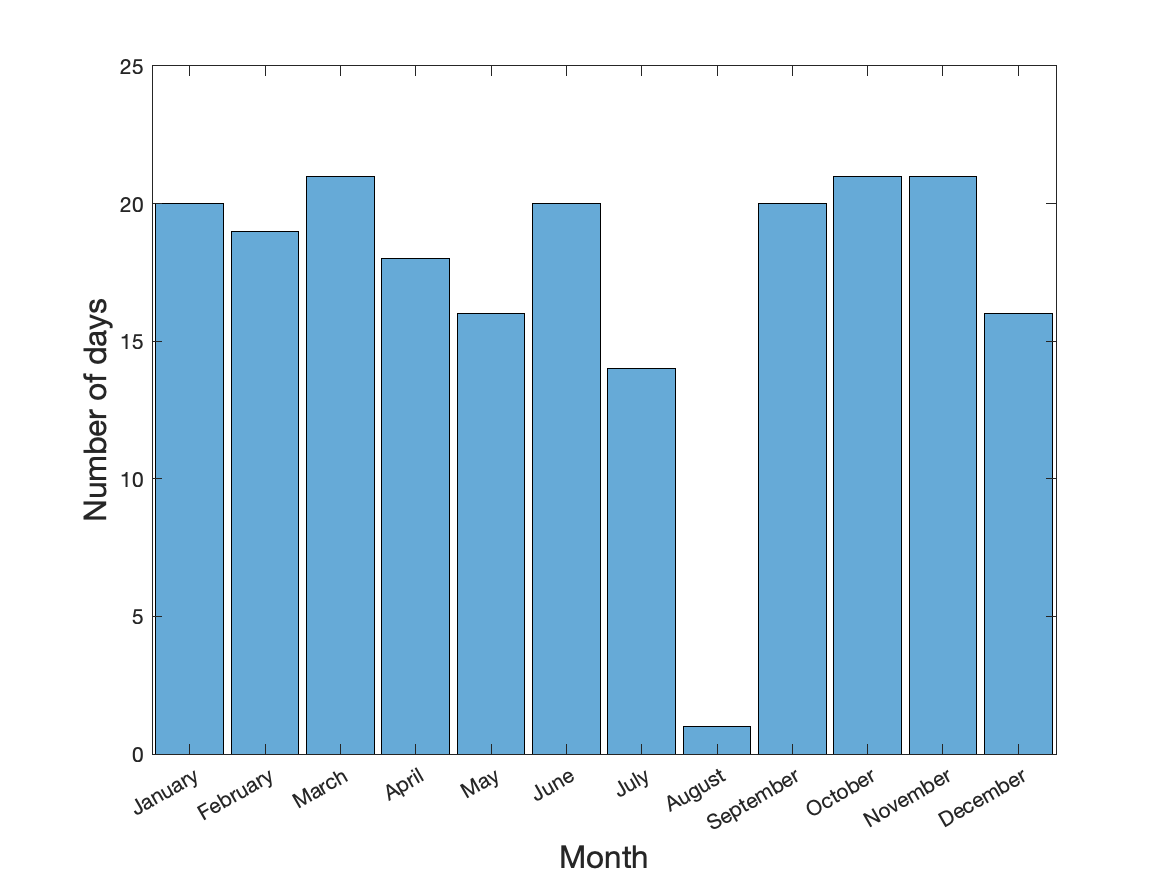}
    \end{minipage}  & 
 \begin{minipage}{.25\textwidth}
 \vspace*{0.02in}   \includegraphics[trim = 20 0 0 0, width=\linewidth, height=29mm]{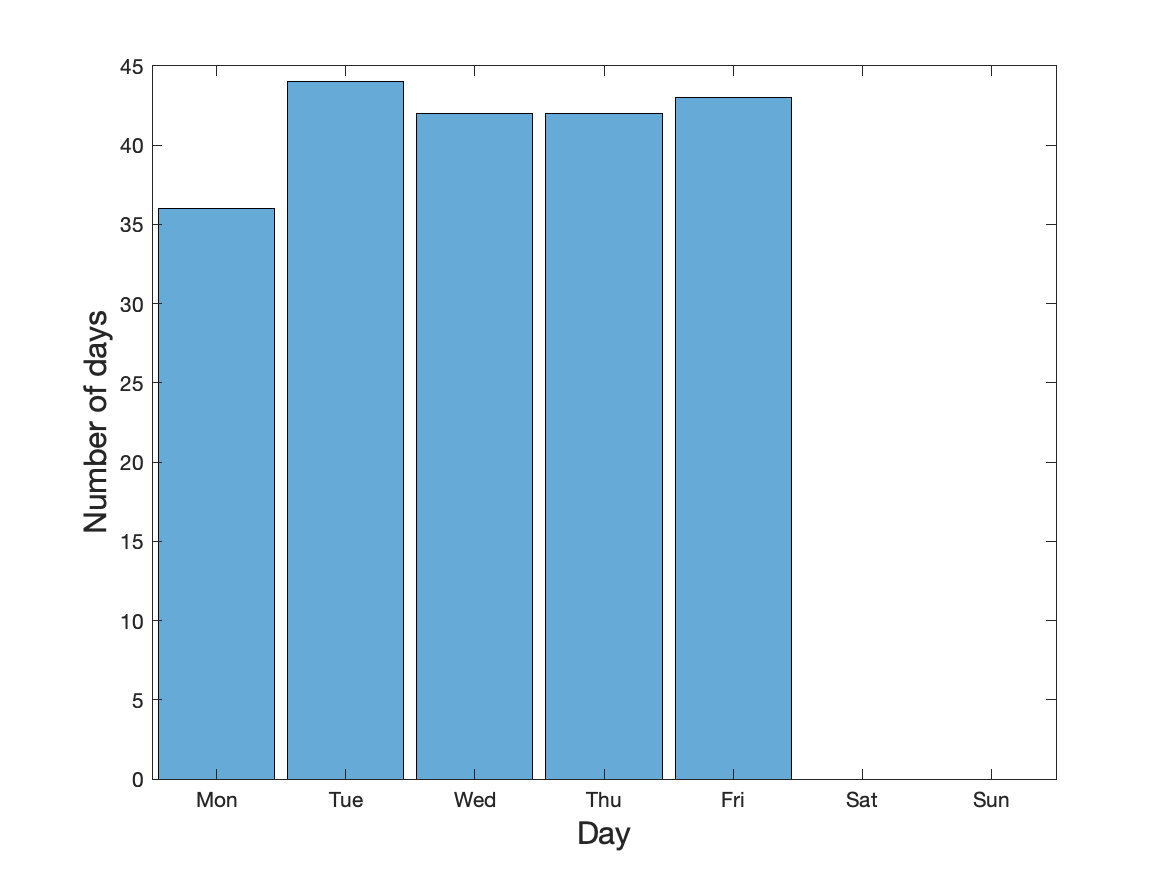}
    \end{minipage} \\
\hline     
6 &
 \begin{minipage}{.3\textwidth}
 \vspace*{0.02in}
  \includegraphics[trim = 20 0 0 0, width=\linewidth, height=29mm]{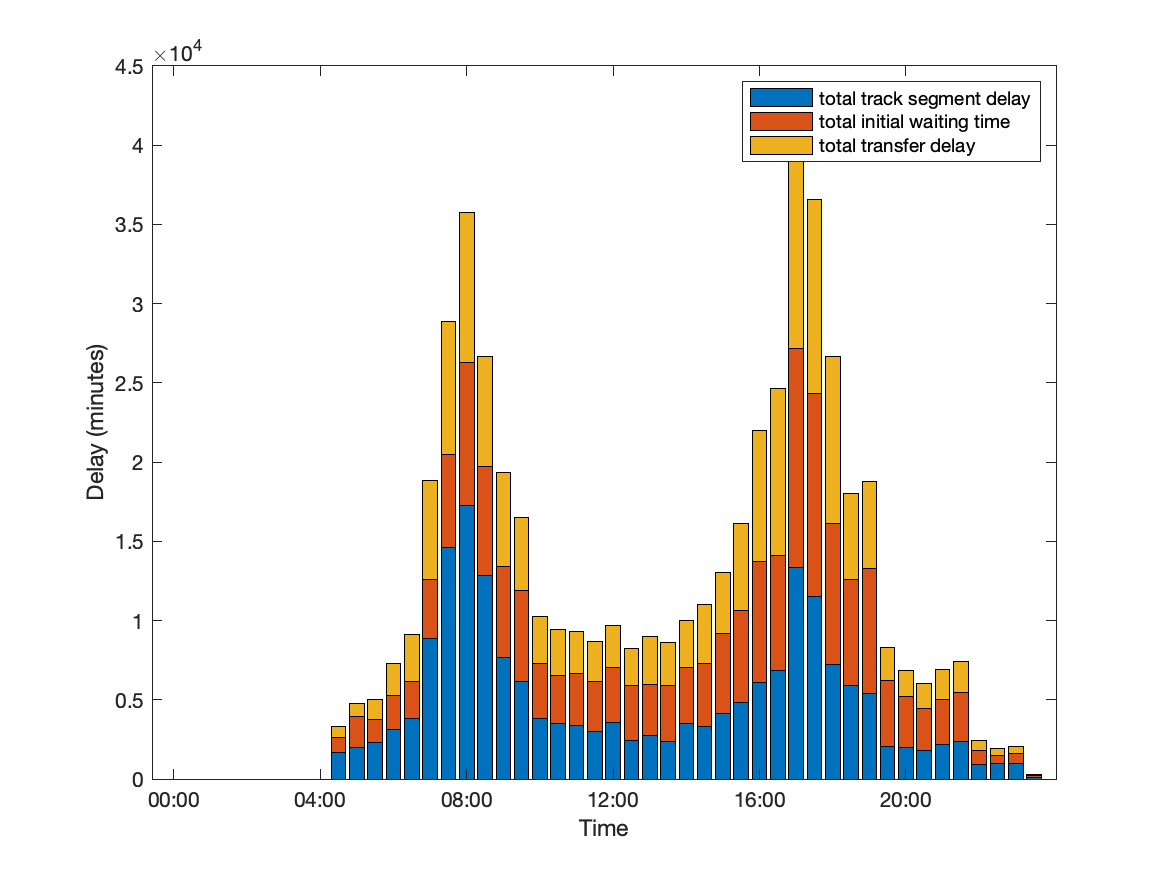}
    \end{minipage} & 
 \begin{minipage}{.25\textwidth}
 \vspace*{0.02in}      \includegraphics[trim = 20 0 0 0, width=\linewidth, height=29mm]{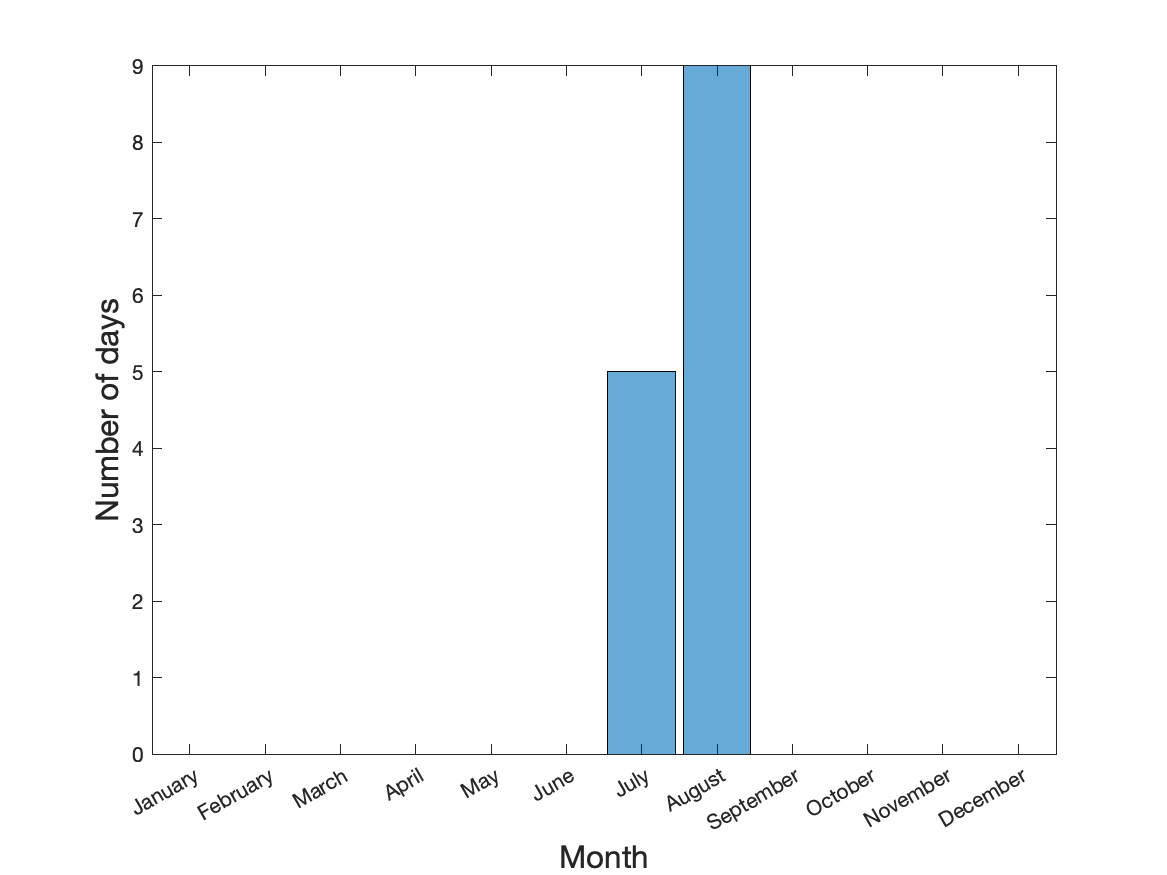}
    \end{minipage}  & 
 \begin{minipage}{.25\textwidth}
 \vspace*{0.02in}   \includegraphics[trim = 20 0 0 0, width=\linewidth, height=29mm]{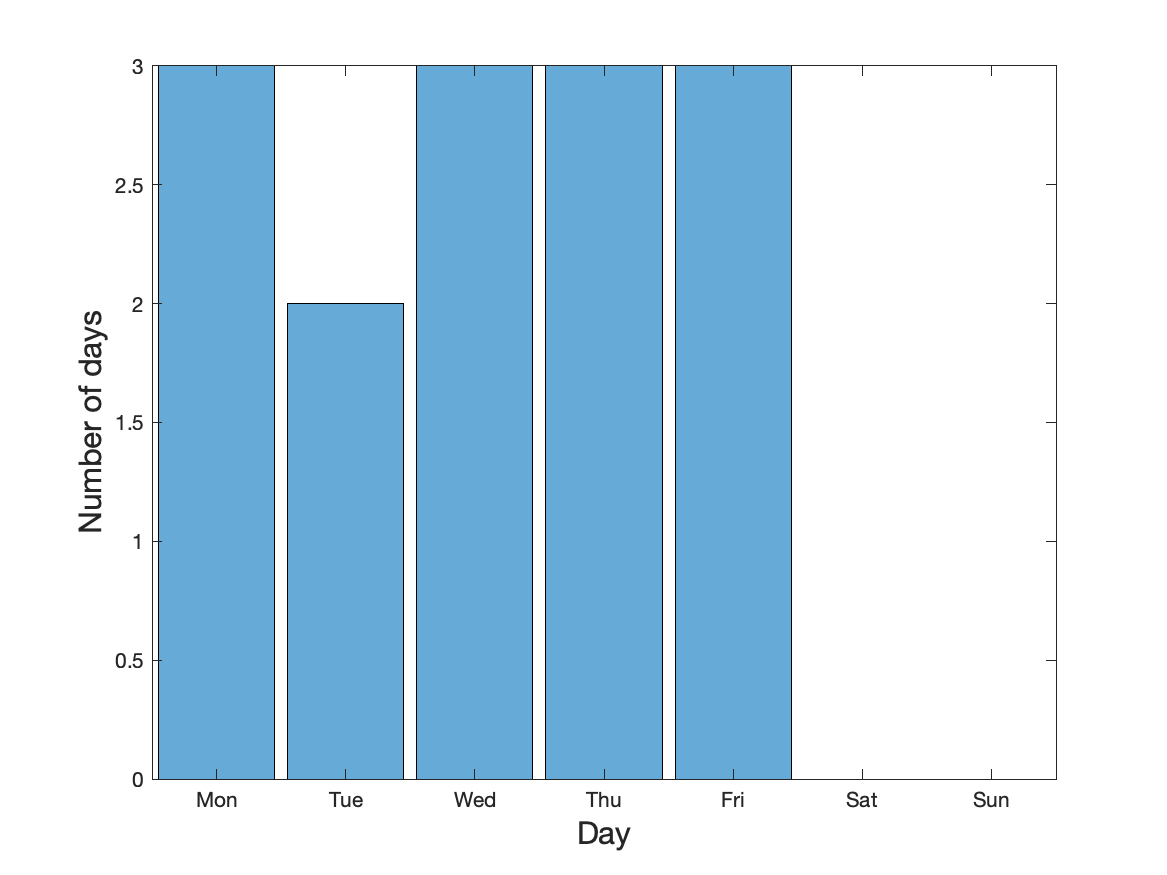}
    \end{minipage} \\

\hline
7 &
 \begin{minipage}{.3\textwidth}
 \vspace*{0.02in}
  \includegraphics[trim = 20 0 0 0, width=\linewidth, height=29mm]{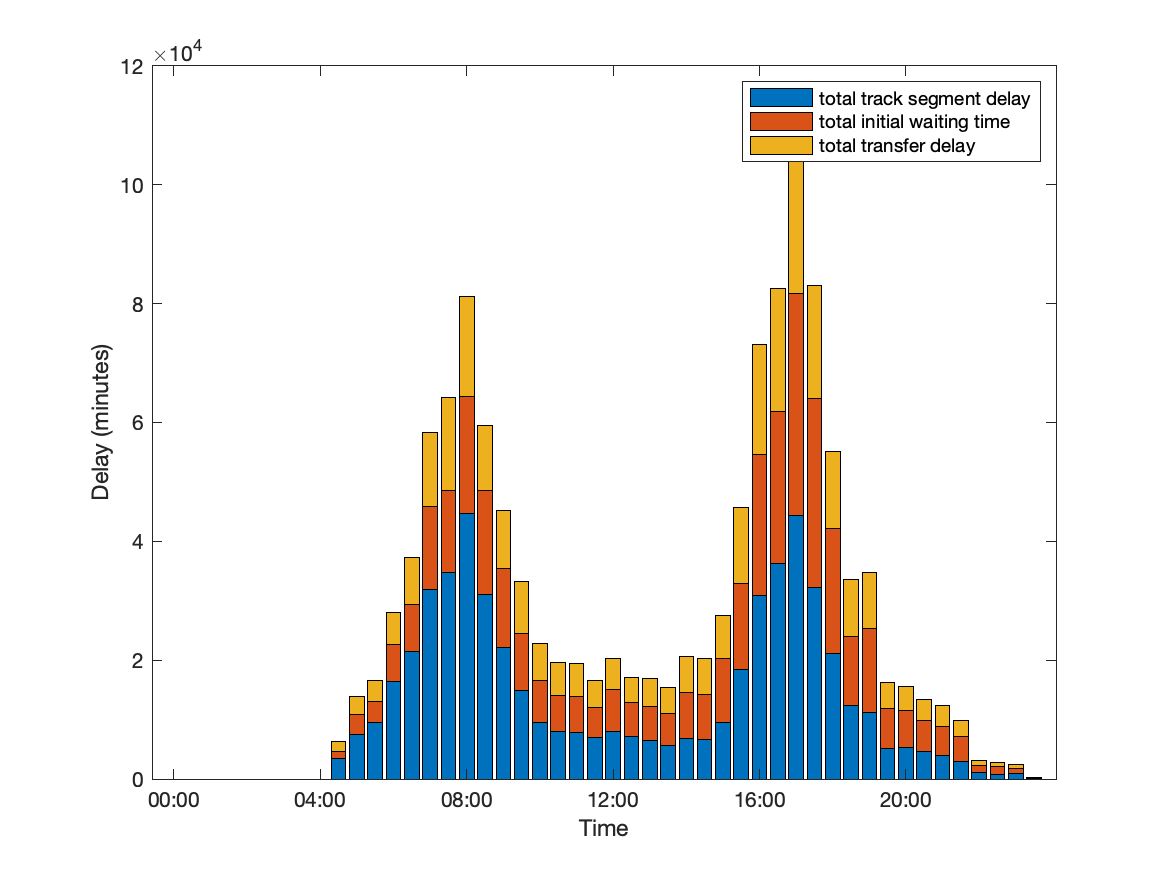}
    \end{minipage} & 
 \begin{minipage}{.25\textwidth}
 \vspace*{0.02in}      \includegraphics[trim = 20 0 0 0, width=\linewidth, height=29mm]{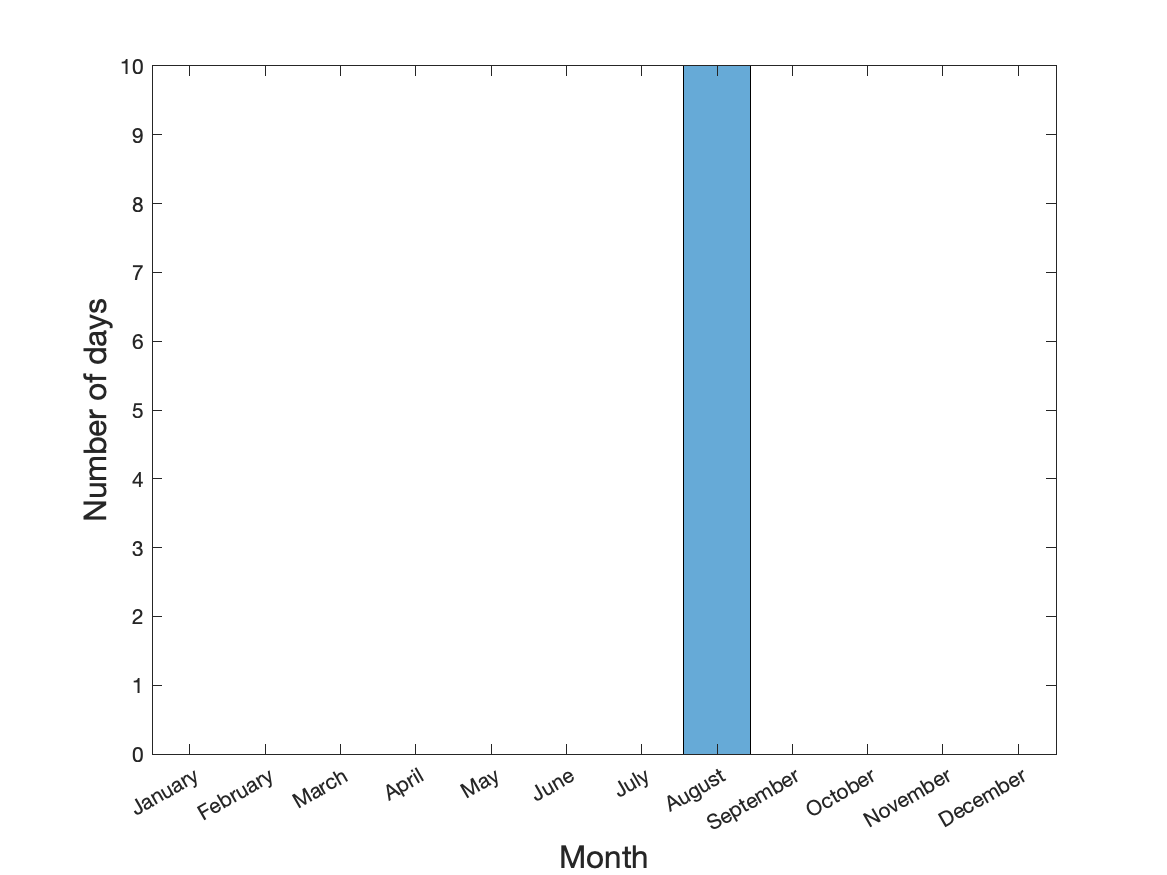}
    \end{minipage}  & 
 \begin{minipage}{.25\textwidth}
 \vspace*{0.02in}   \includegraphics[trim = 20 0 0 0, width=\linewidth, height=29mm]{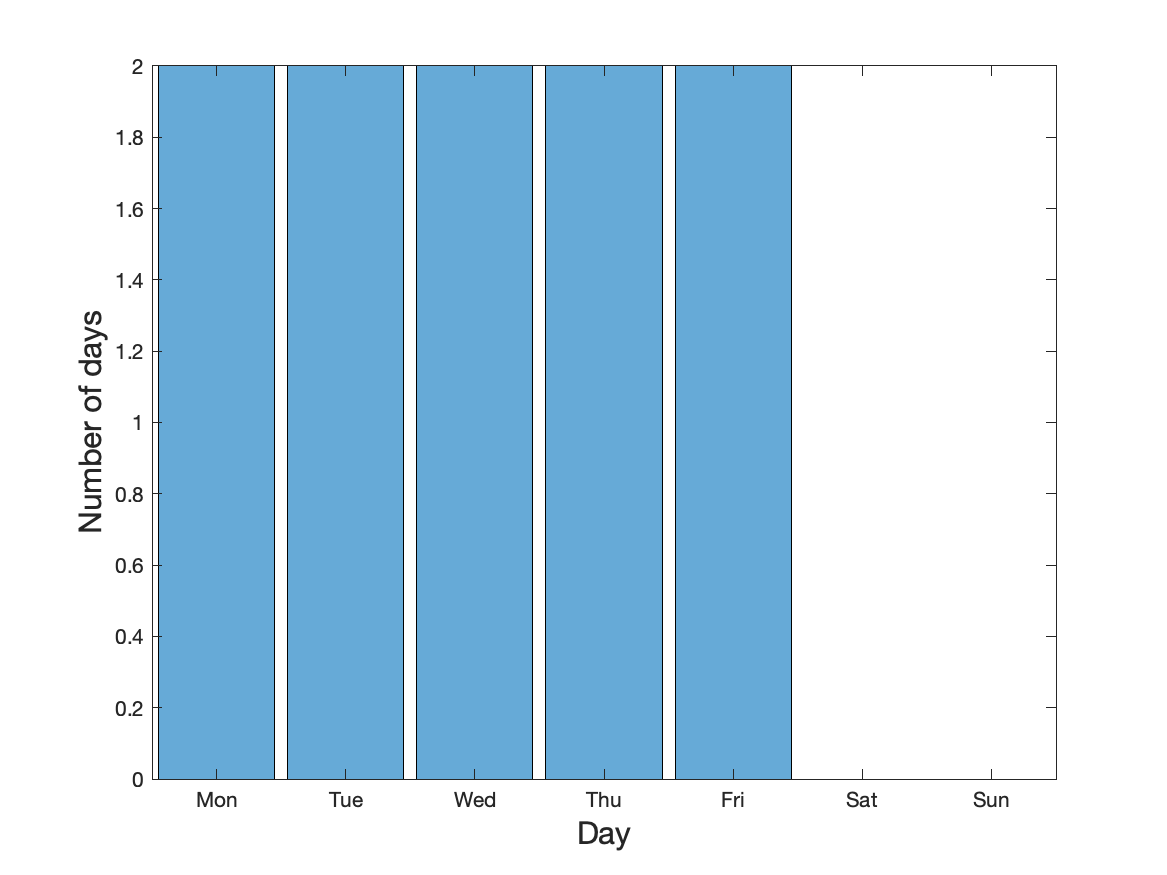}
    \end{minipage} \\
\hline
\end{tabular}
\end{table}

A more in-depth look at each cluster and their corresponding delay distribution of the average passenger delays is given in Table~\ref{classes1}. By looking at the distribution of the days and months in each cluster of the average passenger delay in Table~\ref{classes1}, we make the following observations:
\begin{itemize}
    \item There are distinct weekday and weekend patterns.
    \item For all months, Sundays exhibit distinct patterns which are mainly manifested in classes 2, 3 and 7 with relatively high average passenger delay.
    \item In class 3, it is exclusively Sunday but the outlier Monday and Thursday clustered there corresponds to Christmas, New Year and Thanksgiving day respectively.
    \item Class 5 consists exclusively of Saturday except for a Wednesday which corresponds to a special holiday - 4th of July.
    \item August and December exhibit distinct patterns from other months. 
\end{itemize}

An in-depth look at the  total passenger delay in Table~\ref{classes2} provides insight into the number of passengers affected on different days and months. We make the following observations based on Table~\ref{classes2}:
\begin{itemize}
    \item Weekdays and weekends have distinct patterns.
    \item Weekdays have two distinct peaks in total passenger delay whereas weekends have more stable passenger delay.
    \item August and December have distinct patterns compared to other months.
    \item There is no clear seasonal variance other than these outlier months.
    \item The total passenger delay is higher in the month of August with an additional 10,000 passenger delay minutes. From Table~\ref{classes1}, the delay per passenger for the month of August is also high. This can be attributed to August being one of the busiest month in Washington DC.
    \item Majority of the days belong to classes 2 and 5 which have relatively low number of passengers affected by delay.
\end{itemize}

Even though there is a clear division between the delay for different days and different months, our assumption of setting the number of clusters as 7 for creating distinct classes containing only Mondays, Tuesdays, etc did not pan out. This is mainly because of the month of August which biased the clustering. However, the clustering revealed distinct weekday and weekend patterns for both average and total passenger delay. Saturdays are clustered together with weekdays more often than Sundays. Some weekdays are clustered into the Sunday groups. Our assumption is that this is a special day and we have confirmed this at least for the main holidays such as Christmas, New year and Thanksgiving.
\section{Conclusion}
\label{sec:conclusion}

In this study, we propose a new estimation method to map passenger delay into corresponding network elements. The outputs of our method can aid in measuring network performance for any given origin-destination pair of the public transportation network and in prioritizing measures for improving service robustness. We decompose the delay along a passenger trajectory into its corresponding track segment delay, initial waiting time and transfer delay. We demonstrate the method on one year data from Washington metro network. An initial analysis of the data shows that 14\% of the passengers experience a mean delay of 6 minutes or more in this network when looking across the different days. The average passenger delay is more or less stable throughout the day whereas total passenger delay, which accounts for the number of passengers affected, contains two distinct peaks. The two peaks correspond to the morning peak and evening peak with the evening peak associated with greater passenger delay than the morning peak. The initial waiting time contributes the most to the average passenger delay with 59\% whereas the track segment delay contributes the most to the total passenger delay with 41\% of the total delay being attributed to it. Thus, the track segment delay is more correlated with flows than the initial waiting times. The results of the temporal clustering indicate that there are noticeable differences between weekdays and weekends and recurrent weekly patterns are also manifested. Interestingly, the seasonal variance is mainly evident for the months of August and December, with August particularly standing out.

The estimation approach adopted in this study can be transferred to other locations. It is necessary to ensure that the system of equations is solvable. Further, a different temporal aggregation might be needed depending on the headway between the trains. Even though some seasonal variances could be identified, the month of August biased the clustering approach. Hence, a further avenue of research is to come up with an iterative hierarchical clustering to reveal more detailed daily and seasonal variances. Another direction for future research is to examine if the delay on the road traffic network has any impact on passenger delay. In \cite{lopez2017revealing}, 3D speed maps, similar to our 3D delay maps, are constructed for the urban road network of Amsterdam, which were clustered to reveal 4 so-called consensual patterns to represent 35 days of data. Having a similar network state representation for both road and public transport can aid in understanding the interaction between the two modes. 

The insights gained in this study can aid in developing demand-oriented metro services for different days within different months. Furthermore, it can help in understanding the delay propagation through a network and thus the prediction of such delays. This can help in providing proactive services in case of disruptions, thus improving the service performance by reducing operational costs and shorten waiting times for passengers. 
\hfill
\section{Author Contribution}
The authors confirm contribution to the paper as follows: study conception and design: P. Krishnakumari, O. Cats and H. van Lint; data analysis: P. Krishnakumari; interpretation of results and draft manuscript preparation: P. Krishnakumari, O. Cats. All authors reviewed the results and approved the final version of the manuscript.

\section{Acknowledgement}
This research was supported by the SETA project which is financed by the European Union's Horizon 2020 Research and Innovation program under the grant agreement No 688082. The authors would like to thank Washington Metropolitan Area Transit Authority and in particular Jordan Holt for their valuable cooperation and providing the data that made this study possible.


\printbibliography

\end{document}